\documentclass[12pt,a4paper]{article}
\usepackage{graphicx}
\usepackage{dcolumn}
\usepackage{bm}
\usepackage{hyperref}
\usepackage{braket}
\usepackage{authblk}
\usepackage{float}
\usepackage{mathptmx}  
\usepackage[left=2.75cm,right=2.75cm,top=2.75cm,bottom=2.75cm]{geometry}
\usepackage{fancyhdr}
\usepackage{epstopdf}
\begin{document}

\title{Position measurement-induced collapse states:\\ Proposal of an experiment}
\author[1,2]{ Moncy V. John\thanks{moncyjohn@yahoo.co.uk}}
\author[3]{Kiran Mathew \thanks{kiran007x@yahoo.co.in}}
\affil[1]{ Department of Physics, St. Thomas College, Kozhencherri - 689641, Kerala, India}
\affil[2]{School of Pure and Applied Physics, Mahatma Gandhi University, Kottayam - 686560, Kerala, India}
 \affil[3]{Department of Physics, Pavanatma  College, Murickassery -  685604, Kerala, India.}

\date{\today}

\maketitle

\begin{abstract}
 The quantum mechanical treatment of diffraction of  particles, based on the standard postulates of quantum mechanics and the postulate of existence of quantum trajectories,  leads to the `position measurement-induced collapse' (PMIC) states. An experimental set-up to test these PMIC states is proposed. The apparatus consists of a modified Lloyd's mirror in optics, with two reflectors instead of one. The diffraction patterns for this case predicted by the PMIC formalism are presented. They exhibit  quantum fractal structures in space-time called `quantum carpets', first discovered by Berry (1996). The PMIC formalism in this case closely follows the `boundary bound diffraction' analysed  in a previous work by Tounli, Alverado and Sanz (2019). In addition to obtaining their results, we have shown that the  time evolution of these collapsed states also leads to Fresnel and Fraunhofer diffractions. It is anticipated that the verification of PMIC states by this experiment will help to  better understand collapse of the wave function during quantum measurements. 
\end{abstract}

\section{Introduction}
With the advancement of technology in dealing with single-particle systems,  quantum theory has now entered a new phase, namely that of `quantum measurement'  \cite{kip,wise}. This has  led to new operational interpretations of quantum mechanics. In classical mechanics of a system of particles,   measurement of  position   is of  prime concern. The case of quantum mechanics is not very different either. Heisenberg's thought experiment, in which the position of a particle is measured with an idealised `gamma ray microscope', has been central to its understanding. However, quantum theory lacked a proper operational definition of `position measurement'  for a long time.  Lamb \cite{lamb} has made an  attempt in this direction, observing that the above thought experiment by Heisenberg can only be considered  as a scattering experiment.  Recently operational  approaches to  quantum position measurement have  gained renewed attention.

The question  whether  single-slit diffraction can be treated as quantum measurement of position of a particle was raised  by Marcella \cite{marcella} and was followed up in \cite{rothman,fabbro}.  In a recent paper, we have obtained a satisfactory, affirmative answer  \cite{mvjkm} to the above question using the standard axioms of  quantum mechanics,  together with  nonlocal quantum trajectory representations.  As in \cite{marcella}, we have considered  a particle that passes through a slit at a time $t=t_M$ and assumed that  its wave function  collapses to a rectangular function.  We then expressed this collapsed rectangular wave function (which can be considered   as a superposition of position eigenfunctions)  in terms of the energy eigenfunctions in the relevant  case.  Lastly, unitary time-evolution of the state  for   $t\geq t_M$ is introduced and this results in what we called the position measurement-induced collapse (PMIC) state. The formalism could lead to  a unified quantum description of Fresnel (near-field) and Fraunhofer (far-field) diffractions. The result that a single quantum expression describes both kind of diffractions was claimed to be the first its kind.

   In the present Letter  we  describe the PMIC  states for a particle whose position is measured when it is in an infinite potential well and then suggest an experimental set-up where this formalism can be tested. For this purpose,  a  modified single-slit diffraction arrangement, which can be considered  as a double Lloyd's mirror apparatus, is proposed. Here we have adopted  a purely quantum mechanical approach based on PMIC states that can be verified with experiment.

   The present work has some closely related antecedents in \cite{berry98} and \cite{sanz19}. In the former, Berry has computed the probability density corresponding to a wave function of a particle in a box that evolves from an initial one with a discontinuity at its walls. This work has drawn attention to some unexpected fractal properties of this function.  In a recent paper,  Tounli, Alvarado and Sanz \cite{sanz19} have considered  `boundary-bound diffraction' of a spatially localised matter wave   with   discontinuity at its boundary. Thereafter, the matter wave is considered to evolve freely beyond the initial localisation region, but is frustrated by the presence of hard-wall-type boundaries of a cavity that contains it. Both these works lead to fractal functions in space and time. In the latter paper, the authors have  noted that the development of space-time pattern inside the cavity depends on (1) the shape of the wave function $f(x)$ in the initial localisation region, (2) the mass of the particle considered and (3) the relative extension of the initial state with respect to the total length spanned by the cavity.  However, they did not identify any Fresnel or Fraunhofer patterns in their analysis. Instead, they state that because of the presence of confining boundaries, even in the cases of an increasingly large box length, no Fraunhofer-like diffraction features can ever be observed at any time.  We have  reproduced the fractal patterns obtained by them, which are referred to as `quantum carpets' for the cases mentioned. The revival of wave functions, as found in these papers are also observed. In addition, we  have computed the probability density of particles on the screen placed at various distances from the slit. Contrary to their conclusion regarding diffraction patterns, we  found that Fresnel and Fraunhofer patterns are indeed present in them. The diffraction experiment suggested by us shall help to demonstrate the above fractal and nonfractal features of the wave function at various times and also the predicted revival distances to the screen.

   This paper is organised as follows. The theoretical formulation of PMIC states  is presented in Sec. 2. The  prediction of the probability  patterns that may be obtained for the  case of a particle in an infinite potential well  for various times is made in Sec. 3.  In Sec. 4, the experimental arrangement that can produce these patterns is described. The last section comprises a discussion.

\section{PMIC states} \label{sec:pmic}
In this section, a general formulation of the PMIC states  that closely follows the discussion in  \cite{berry98,sanz19} is made. Since our interest is  to apply this  to an actual experiment of diffraction of matter waves, we continue to view it as  position measurement-induced collapse  state,  defined  using  the  postulate of reduction of the wave function in quantum mechanics \cite{cohen}. According to this postulate, if a measurement of an observable ${\cal A}$ on a system  in the state $\ket{{ \psi}}$  has yielded the result $\alpha_0$ to within an accuracy $\Delta \alpha$, the state of the system immediately after the measurement is described by

\begin{equation}
\ket{\psi^{\alpha_0,\Delta \alpha}}=\frac{1}{\sqrt{\bra{{ \psi}}P_{\Delta \alpha}(\alpha_0)\ket {{ \psi}}}}P_{\Delta \alpha}(\alpha_0)\ket{{ \psi}},
\end{equation}
     with the projection operator

\begin{equation}
P_{\Delta \alpha}(\alpha_0)=\int_{\alpha_0-\Delta \alpha /2}^{\alpha_0+\Delta \alpha /2}d\alpha \ket{{\alpha}} \bra{{\alpha}}.
\end{equation}
Here $\{ \ket{{\alpha}} \}$ is the set of eigenstates of ${\cal A}$ that serves as a complete orthonormal basis. Let us consider the collapse of the wave function of a particle in one-dimension with coordinate $y$, under the measurement ${\cal A}$.
With $\braket{{\alpha}|{ \psi}}\equiv { c}(\alpha)$ and $\braket{y|{\alpha}}\equiv v({\alpha},y)$, one can write the collapsed wave function $\ket{\psi^{\alpha_0,\Delta \alpha}}$ in the position representation as

\begin{equation}
\psi_y^{\alpha_0,\Delta \alpha}(y)=\frac{1}{\sqrt{\int_{\alpha_0-\Delta \alpha /2}^{\alpha_0+\Delta \alpha /2}d\alpha \mid { c}(\alpha) \mid^2}}   \int_{\alpha_0-\Delta \alpha /2}^{\alpha_0+\Delta \alpha /2}d\alpha \; { c}(\alpha)\; v({\alpha},y). \label{eq:pmic1}
\end{equation}

Now consider the case where ${\cal A}$ is the  position operator for the particle, representing position measurement.  Let us denote the eigenvalues  and eigenvectors   in this case as $y^{\prime}$ and $\ket{y^{\prime}}$, respectively. Also let  the measurement of position give the value of $y$ in the interval $[y_0-a/2,y_0+a/2]$; i.e., with an accuracy $ \Delta y \equiv a$.  Then the above equation can be rewritten with $\alpha=y^{\prime}$, $\alpha_0=y_0$ and $\Delta \alpha =a$. We also have $ { c}(y^{\prime})\equiv \braket{y^{\prime}|{ \psi}} = { \psi}(y^{\prime})$  and $v({y^{\prime}},y)\equiv  \braket{y|y^{\prime}} = \delta(y-y^{\prime})$. The Dirac delta function in the latter expression is the position eigenstate in the position representation. Eq. (\ref{eq:pmic1}) now becomes

\begin{equation}
\psi_y^{y_0,a}(y)=\frac{1}{\sqrt{\int_{y_0-a /2}^{y_0+a /2}dy^{\prime} \mid { \psi}(y^{\prime}) \mid^2}}   \int_{y_0-a /2}^{y_0+a /2}dy^{\prime} \; { \psi}(y^{\prime})\; \delta(y-y^{\prime}). \label{eq:lin_sup_pos1}
\end{equation}

 In \cite{mvjkm},  the PMIC states is defined using this  expression for  collapsed states. First assume that  immediately after the above collapse, the particle is in a potential $V$, where the eigenstates of energy are $\ket{u_n}$ and the corresponding position space energy eigenfunctions are  $u_n(y)\equiv \braket{y|u_n}$.  We can use  the closure representation of Dirac delta function \cite{arfkenp264}

\begin{equation}
\sum_{n=0}^{\infty} u_n^{\star}(y^{\prime}) u_n(y) = \delta (y-y^{\prime}), \label{eq:closure}
\end{equation}
to represent the position eigenstate of the particle while it is detected at the point $y^{\prime}$ at $t=t_M$. This helps  to expand the collapsed wave function  (\ref{eq:lin_sup_pos1})  as an infinite series,  given by

\begin{equation}
\psi_y^{y_0,a}(y)=\frac{1}{\sqrt{\int_{y_0-a /2}^{y_0+a /2}dy^{\prime} \mid { \psi}(y^{\prime}) \mid^2}}   \int_{y_0-a /2}^{y_0+a /2}dy^{\prime} \; { \psi}(y^{\prime})\; \sum_{n=0}^{\infty} u_n^{\star}(y^{\prime}) u_n(y). \label{eq:lin_sup_pos2}
\end{equation}
The base functions are chosen  to be the eigenfunctions of the Hamiltonian operator to aid the introduction of  unitary time-evolution of the system.

Let the above wave function be denoted as $ \psi_y^{y_0,a}(y) \equiv \Psi_y^{y_0,a}(y,t_M) $ where $t_M$ is the time at which the measurement is made. Now one can   introduce the time-evolution of the wave function of the particle to obtain the PMIC states for $t\geq t_M$ as 

\begin{equation}
\Psi_y^{y_0,a}(y,t)= \frac{1}{\sqrt{\int_{y_0-a /2}^{y_0+a /2}dy^{\prime} \mid { \psi}(y^{\prime}) \mid^2}}   \sum_{n=0}^{N}  \left[ \int_{y_0-a/2}^{y_0+a/2}dy^{\prime}\psi (y^{\prime})\;u_n^{\star}(y^{\prime}) \right]u_n(y)e^{-iE_n(t-t_M)/\hbar},  \label{eq:slit_wavefn2}
\end{equation}
where  the upper limit $N \rightarrow \infty$ may be taken. Here $E_n$ are the energy eigenvalues of the particle when it is in this potential.

Let us now assume for simplicity that the wave function before collapse  $\braket{y|{ \psi}}\equiv { \psi}(y)$  is a constant over the small interval of width $a$. Then the collapsed wave function $\braket{y|\psi^{y_0,a}} \equiv \psi^{y_0,a}(y)$ in the above equation  can be considered as a rectangular wave function at the instant of collapse, given by  

\begin{eqnarray}
\psi_y^{y_0,a}  (y)  =
\left\{
\begin{array}{lr}
\frac{1}{\sqrt{a}}  &\qquad y_0-a/2\leq y\leq y_0+a/2\\
0& \qquad \hbox{for other values of $y$.}
\end{array}
\right.
\label{eq:slit_wavefn1}
\end{eqnarray}
 In this case, the PMIC wave function becomes

\begin{equation}
\Psi_y^{y_0,a}(y,t)= \frac{1}{\sqrt{a}}   \sum_{n=0}^{N}  \left[ \int_{y_0-a/2}^{y_0+a/2}dy^{\prime} \; u_n^{\star}(y^{\prime}) \right]u_n(y)e^{-iE_n(t-t_M)/\hbar}.  \label{eq:slit_wavefn3}
\end{equation}

Now consider that the potential $V$ experienced by the particle after collapse is  an infinite potential well, with $V=0$ in the interval $-L/2 \leq y \leq L/2$ and $V=\infty$ outside. Let the measurement, made  at time $t_M$,   give the particle's  position as  $y_0$ with an accuracy $a$, where $y_0$ is contained inside the interval $[-L/2,L/2]$.   Following the discussion above, one can express the above PMIC wave function   in terms of the eigenstates of energy of the particle in the potential well by choosing

\begin{equation}
u_n(y)=A \sin \left[ \frac{n\pi}{L}\left(y+\frac{L}{2} \right) \right], \label{eq:u_n}
\end{equation}
and

\begin{equation}
E_n=\frac{n^2\pi^2\hbar^2}{2mL^2}\label{eq:e_n}
\end{equation}
in Eq. (\ref{eq:slit_wavefn3}), with $n=1,2,3,..$.   We have plotted $|\Psi_y^{y_0,a}(y,t)|^2$ against $y$ using    Eq. (\ref{eq:slit_wavefn3}),  for various fixed values of $t$, with $t_M=0$. We also chose $\hbar /m=1$ and $L=1$. (The fundamental period, which is called the revival time, is then $T=4/\pi$.) The results are discussed in the next section. 

Such expansions  as that in  Eq. (\ref{eq:slit_wavefn1}) was  performed in \cite{berry98,sanz19} to obtain surprising fractal patterns called quantum carpets.  In the next section, we reproduce their results. The spread in rectangular function for $t>0$ shows several interesting features.  Careful analysis also reveals that these patterns contain Fresnel and Fraunhofer diffraction patterns.

\section{Prediction of patterns}  \label{sec:inf_pot_well}
\begin{figure}[!b] 
  \resizebox {1 \textwidth} {0.4 \textheight }
{\includegraphics {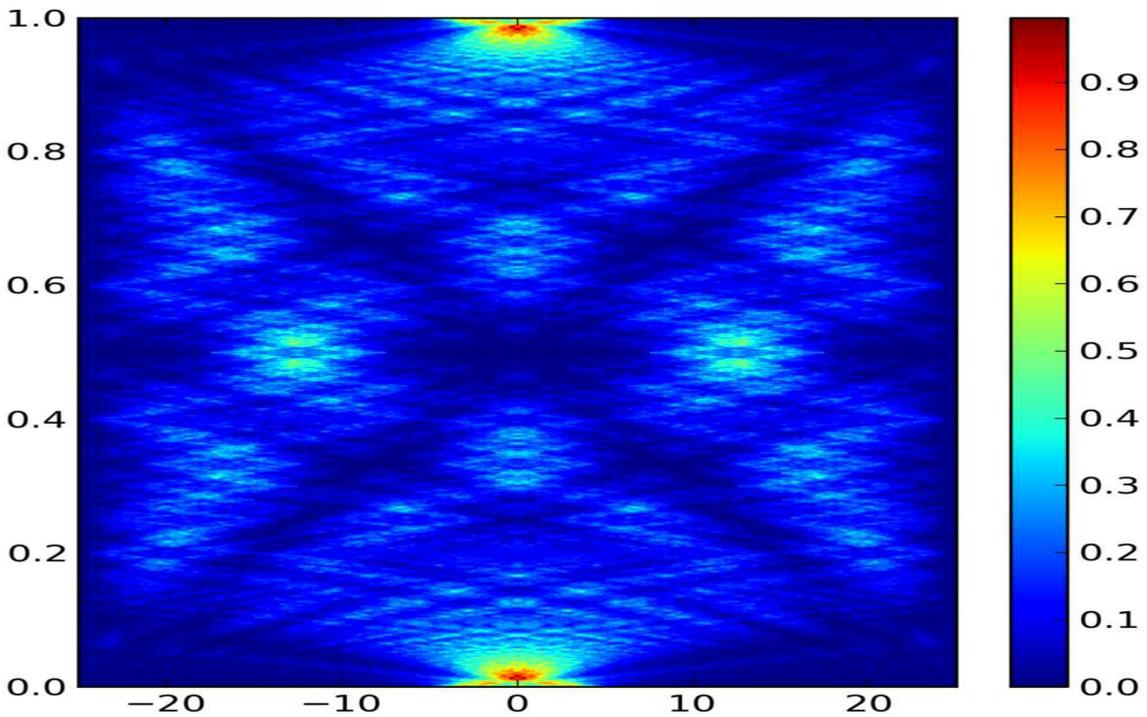}} 
\caption{Quantum carpet }  \label{fig:carpet}
  \end{figure}

The PMIC patterns described by Eq. (\ref{eq:slit_wavefn3}) for the case given in Eq. (\ref{eq:u_n}) and (\ref{eq:e_n}) are presented here with $t_M=0$. First, we  plot the quantum carpet in space-time  for this case of rectangular wave function, with $L=50$, $m/\hbar =1$, $y_0=0$, $a=10$ and $N=500$. The obtained pattern,  shown in Fig. \ref{fig:carpet}, is very similar to the one  in \cite{sanz19}.

\begin{figure}[!b]
\resizebox {0.5 \textwidth} {0.25 \textheight }{\includegraphics {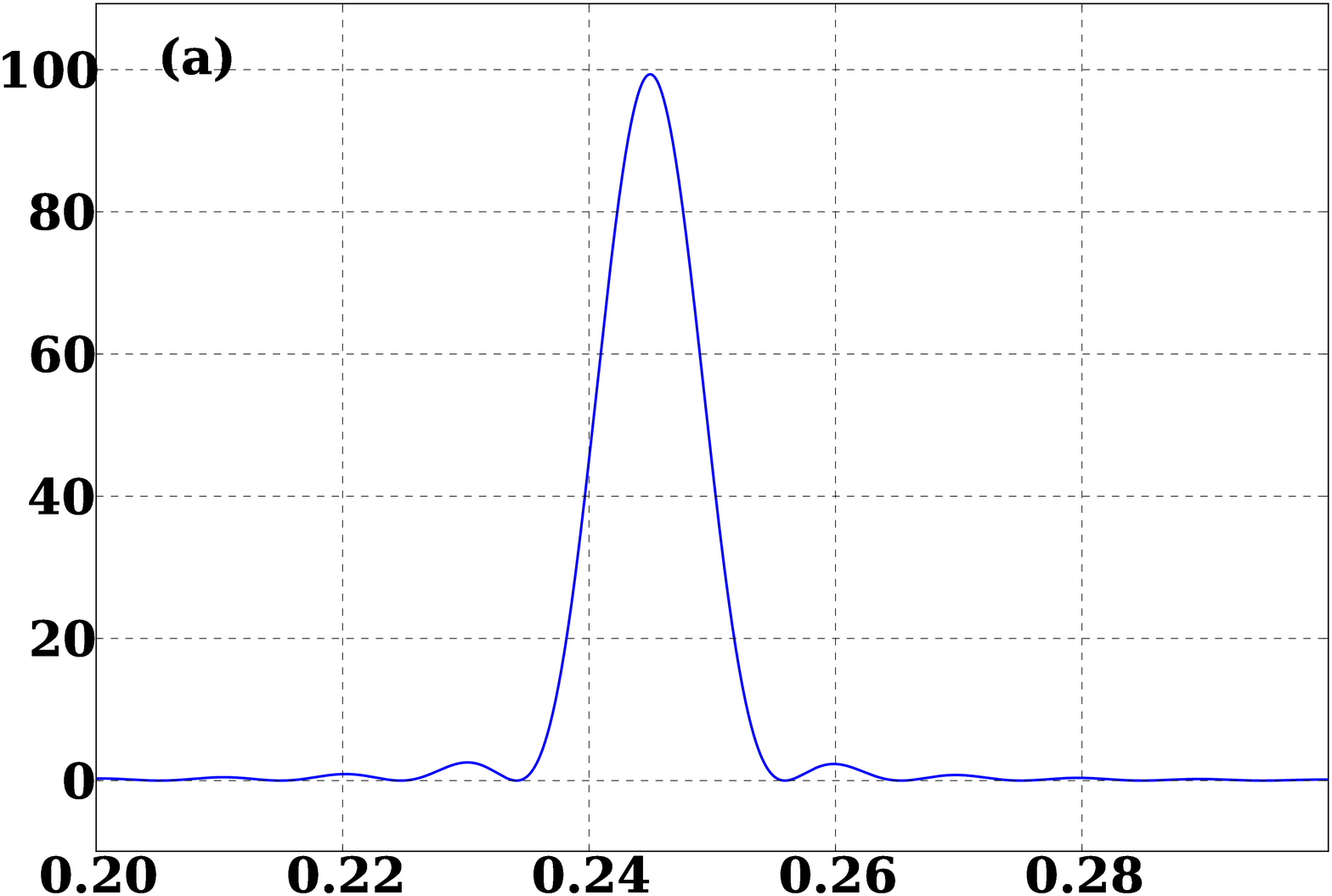}} ,\resizebox {0.5 \textwidth} {0.25 \textheight }{\includegraphics {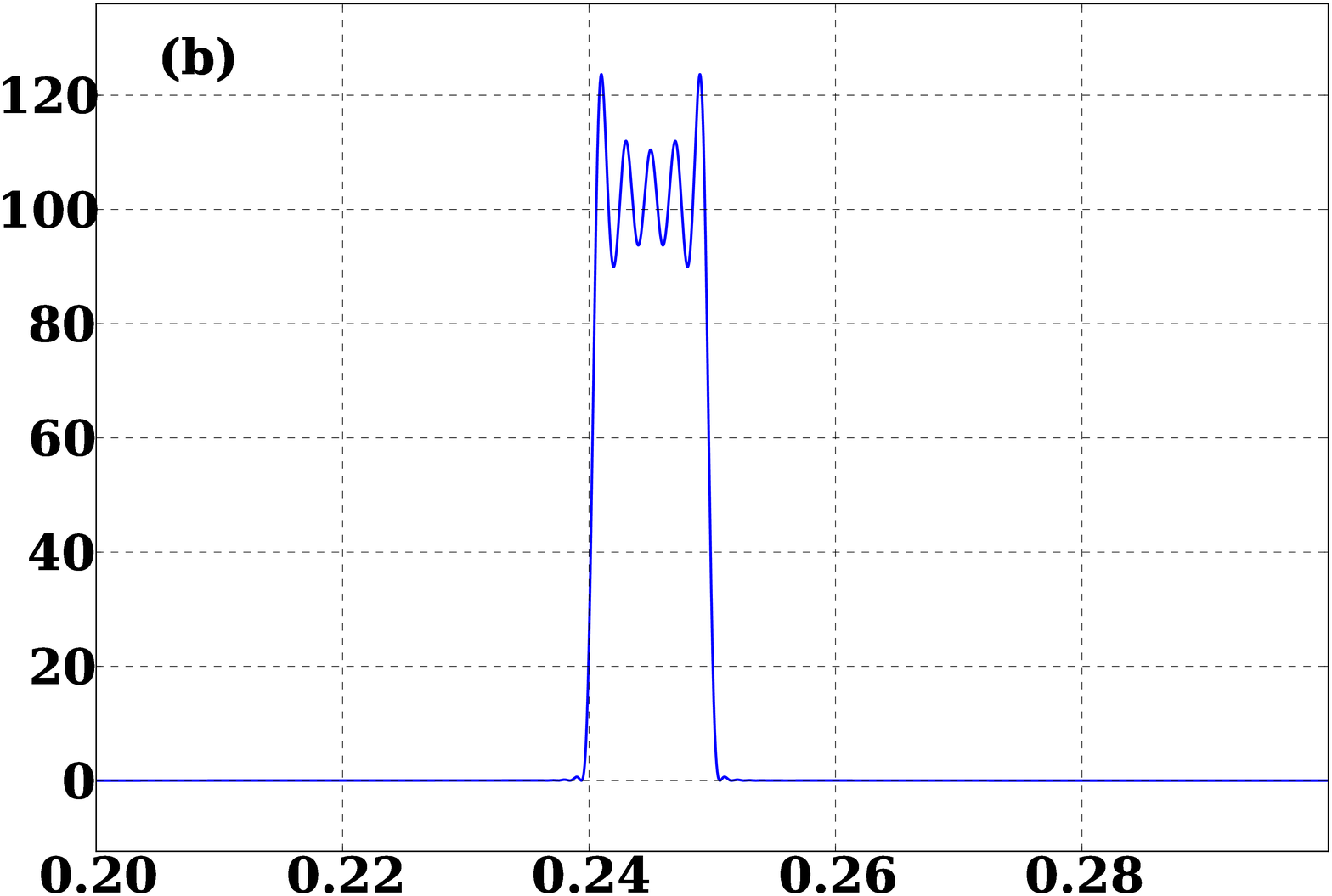}} ,\resizebox {0.5 \textwidth} {0.25 \textheight }{\includegraphics {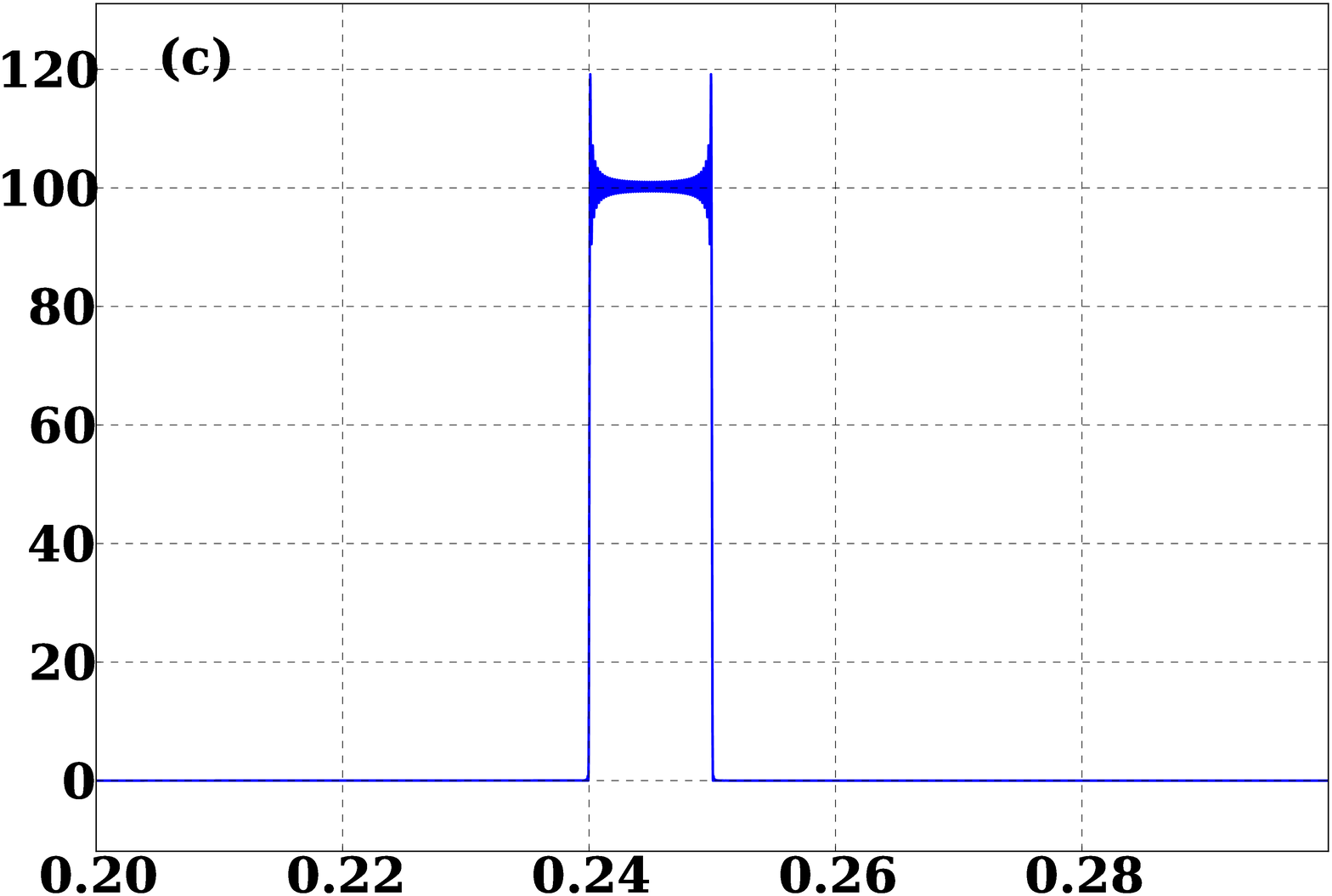}} ,\resizebox {0.5 \textwidth} {0.25 \textheight }{\includegraphics {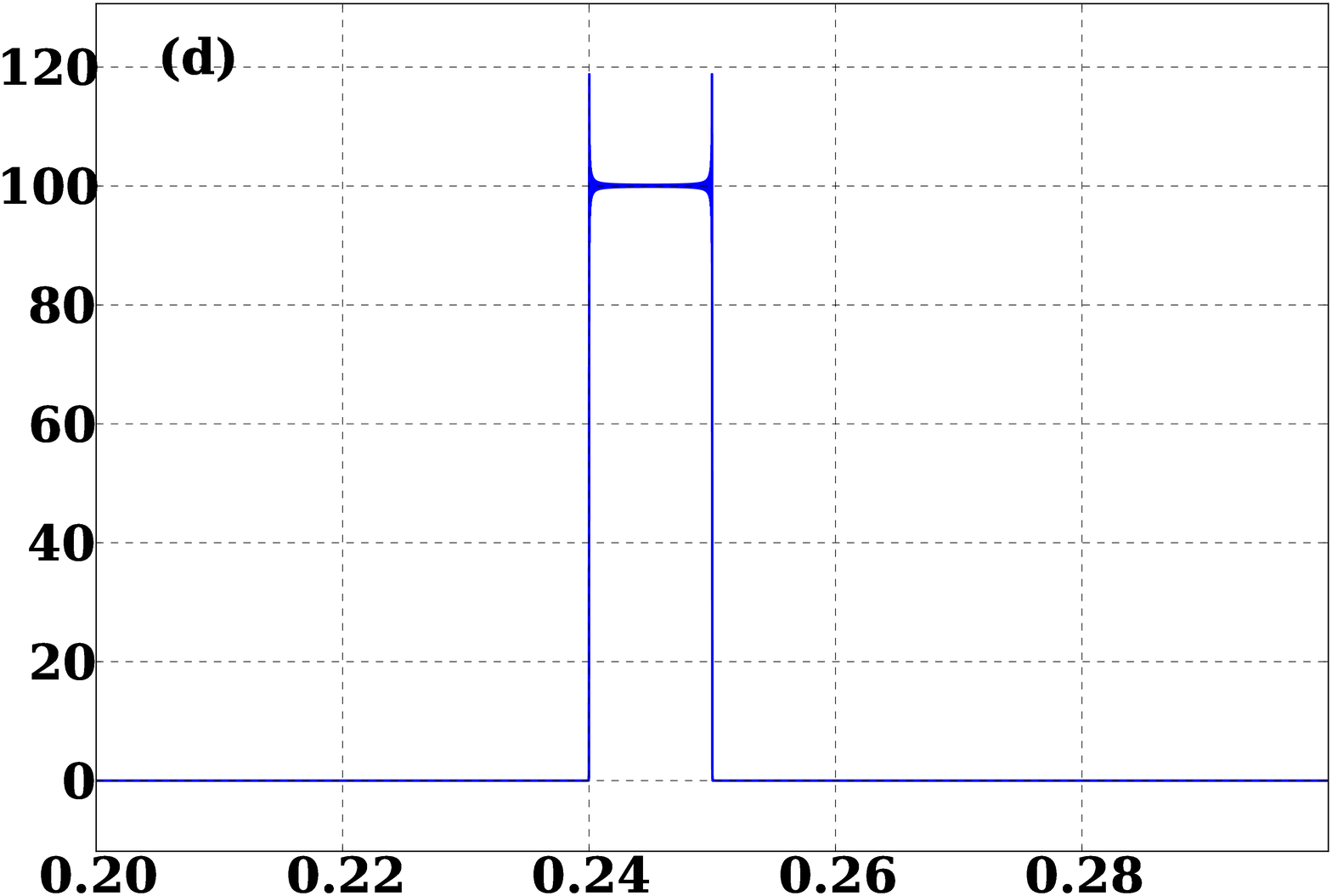}} 
\caption{ PMIC wave function plotted for time $t=0$, with the values of (a) $N=100$ (b) $N=1000$, (c) $N=10000$ (d) $N=50000$}  \label{fig:varyN}
\end{figure}

In an experiment, what one observes is the distribution of particles at some fixed values of $t$. We have plotted such distributions   for various fixed values of $t$, and for different values of $N$, $y_0$ and $a$.  To begin with, we chose $L=1$, $y_0=0.245$ and $a=0.01$.   For comparison, the patterns obtained for $N=100$, $N=1000$, $N=10000$ and $N=50000$   are given in Fig. \ref{fig:varyN} (a)-(d). An almost exact (by eye) rectangular function   is reproduced  with $N=50,000$ and in all the remaining cases discussed below, we have taken this value of $N$ for calculations.

\begin{figure}[!b]
\resizebox {0.33 \textwidth} {0.25 \textheight }{\includegraphics {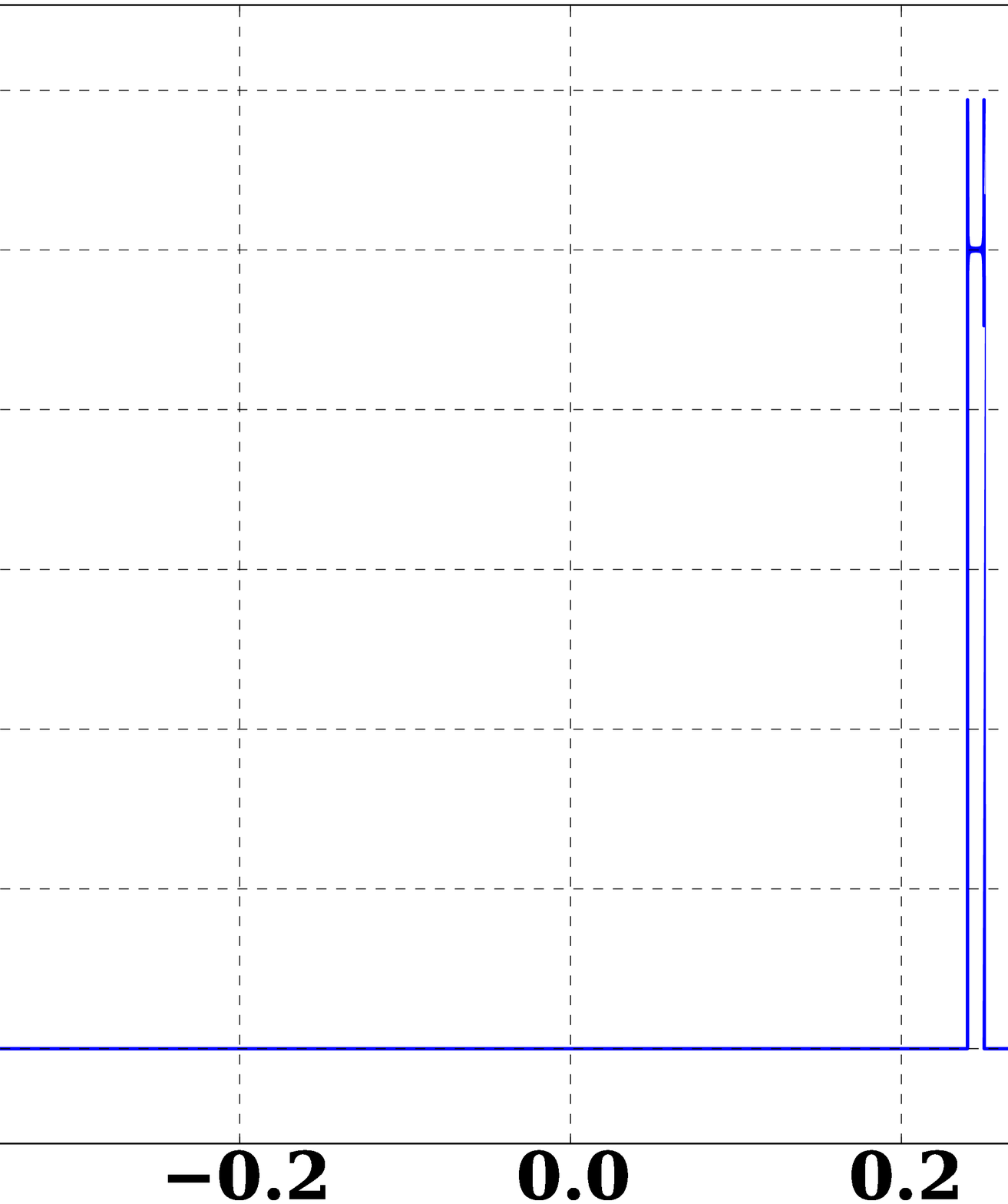}},\resizebox {0.33 \textwidth} {0.25 \textheight }{\includegraphics {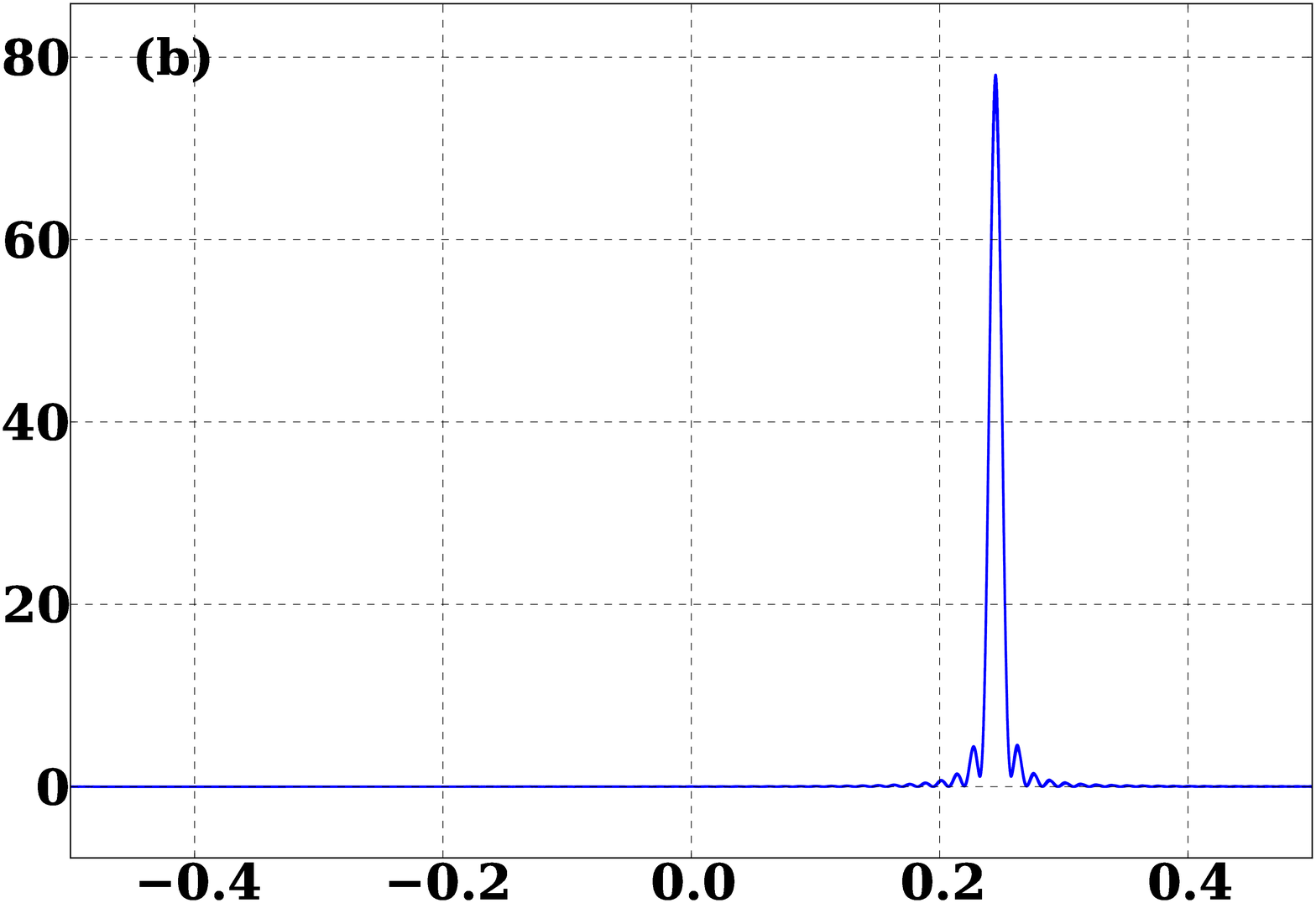}},\resizebox {0.33 \textwidth} {0.25 \textheight }{\includegraphics {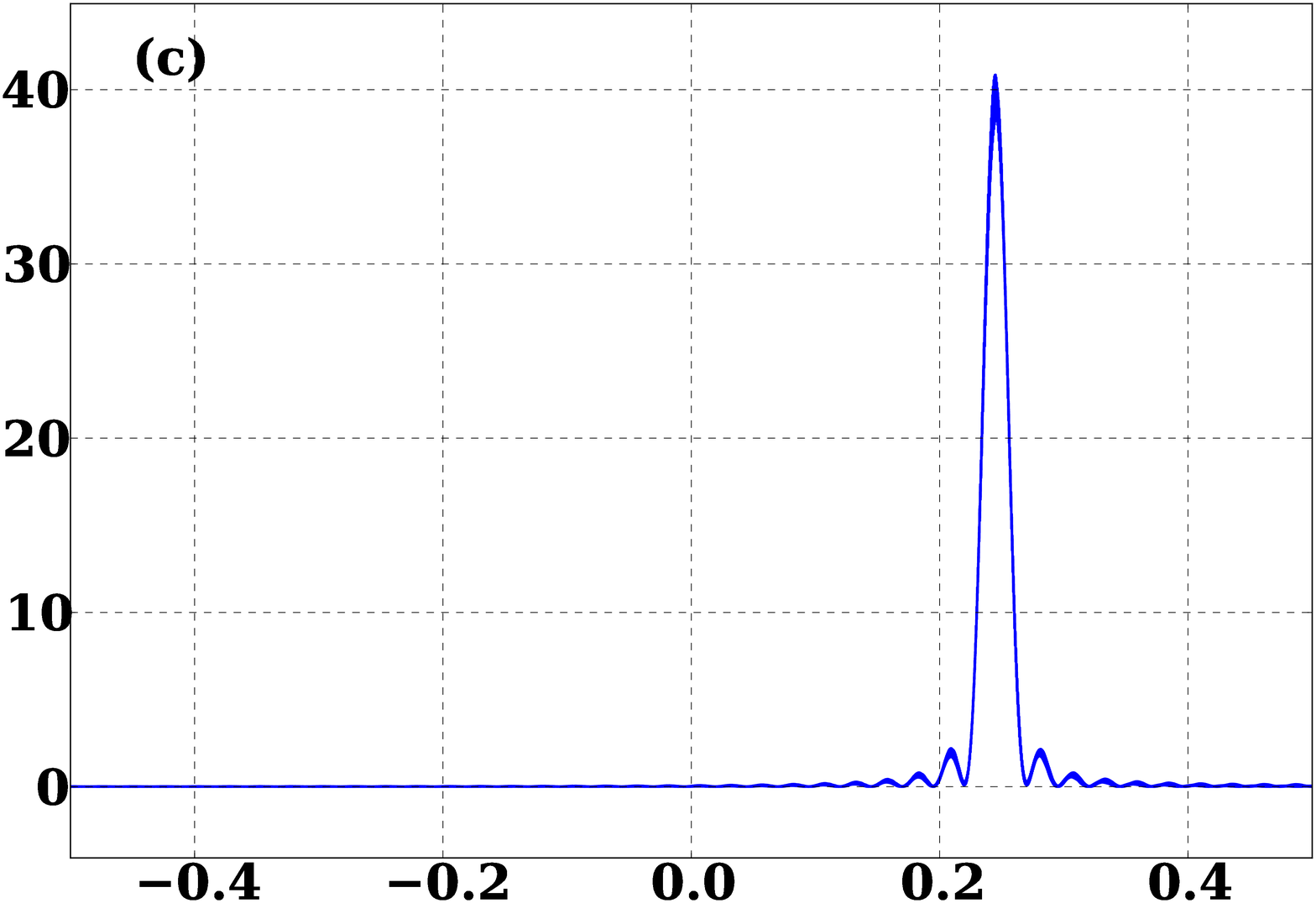}} 
\caption{ PMIC wave function plotted for time $N=50000$, with the values of (a) $t=0$ (b) $t=2\times 10^{-5}$, (c) $t=4\times 10^{-5}$}  \label{fig:varyT_10_5}
\end{figure}

Next, it may be noted that though the rectangular wave function can be obtained for $t=0$ [Fig. \ref{fig:varyT_10_5}(a), same as Fig. \ref{fig:varyN}(d)], 
 a slight variation in $t$  changes this to a Fresnel-type pattern. Here we show such a pattern obtained  at $t=2\times 10^{-5}$ units in Fig. \ref{fig:varyT_10_5} (b). Further, at a time $t=4\times 10^{-5}$ units, it appears almost similar to that of Fraunhofer diffraction, as shown in Fig. \ref{fig:varyT_10_5} (c).

\begin{figure}[!b]
\resizebox {0.5 \textwidth} {0.25 \textheight }{\includegraphics {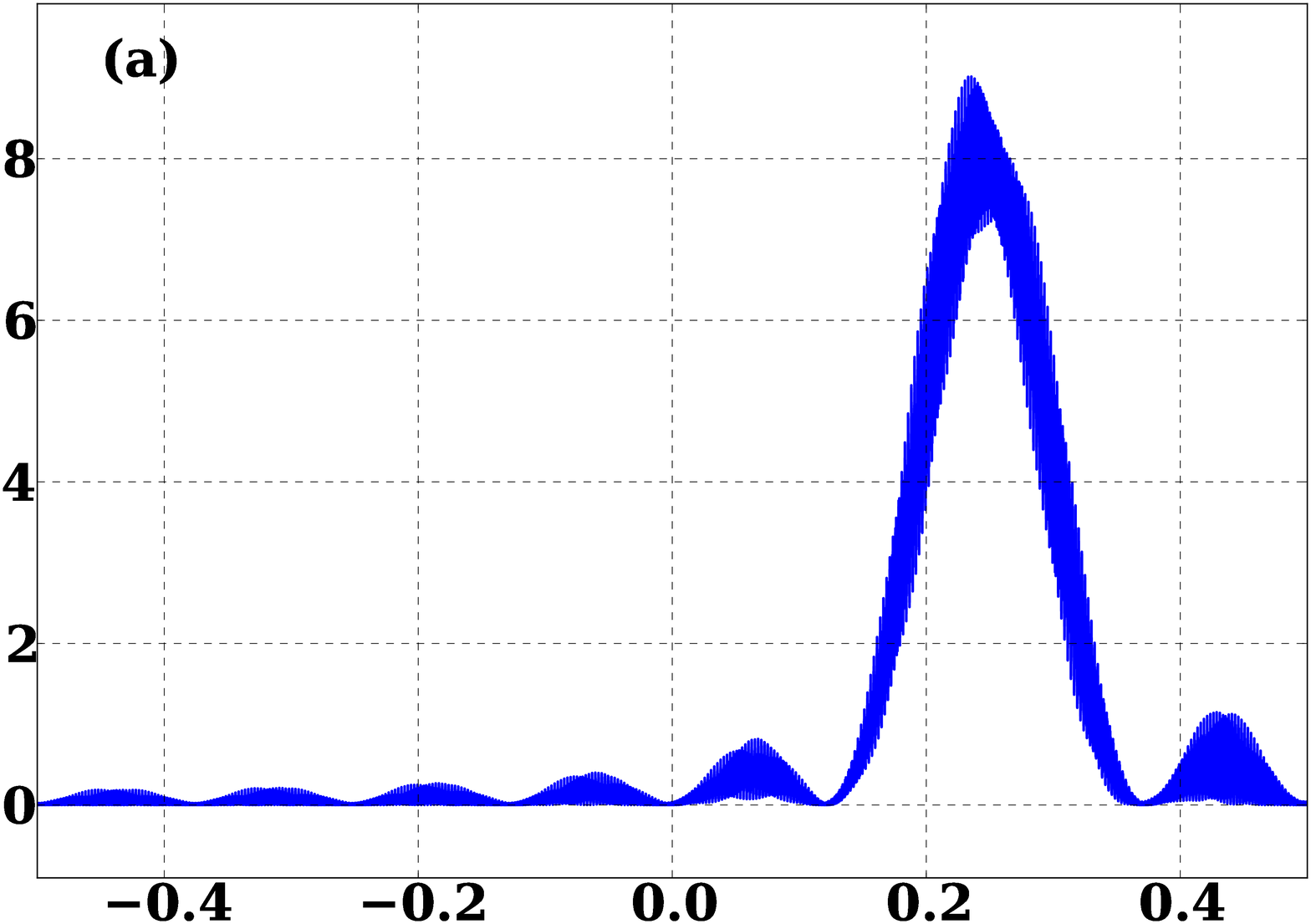}} , \resizebox {0.5 \textwidth} {0.25 \textheight }{\includegraphics {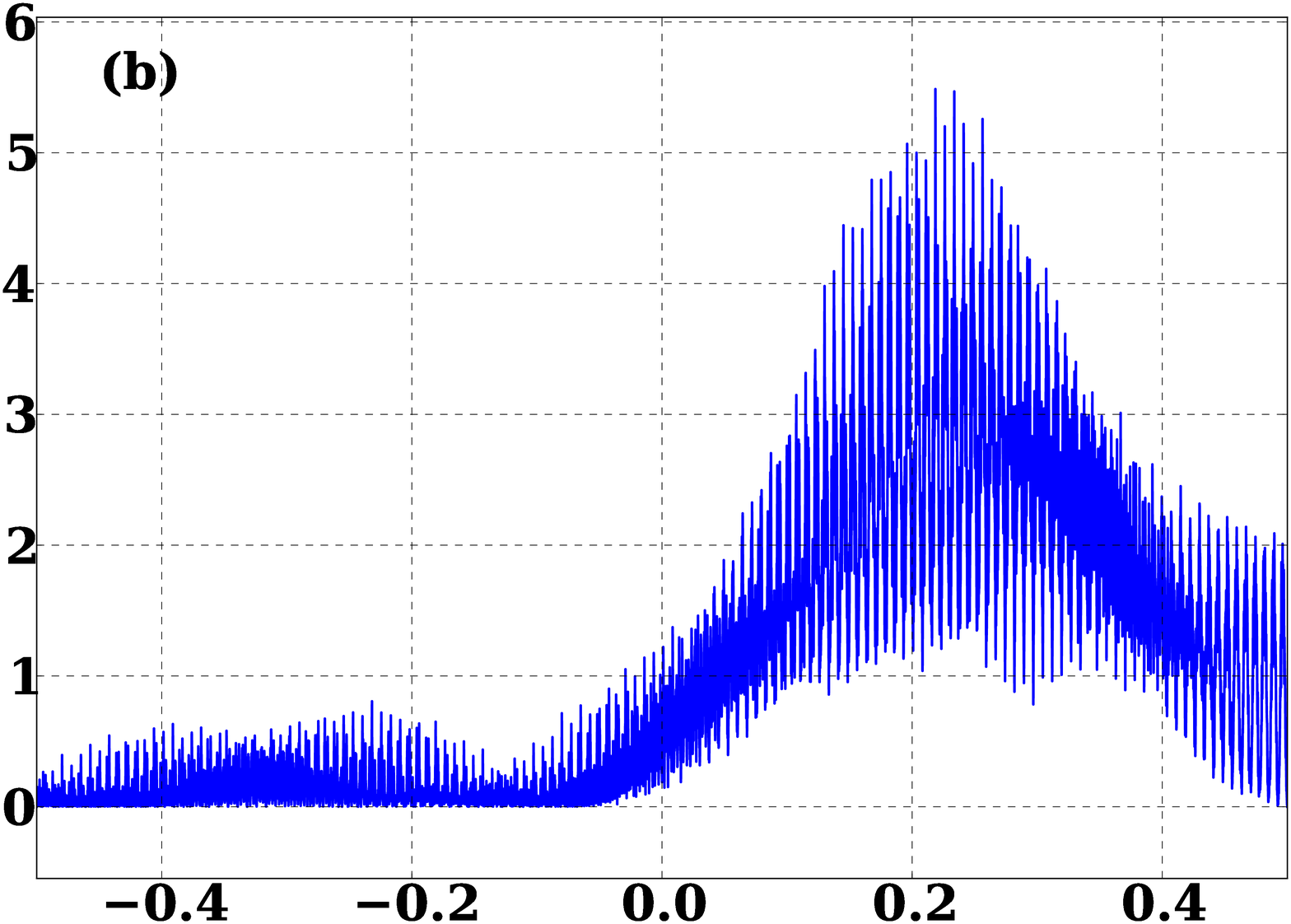}} 
\caption{ PMIC wave function plotted for time $N=50000$, with the values of (a) $t=2\times 10^{-4}$ (b)  $t=6\times 10^{-4}$}  \label{fig:varyT_10_4}
\end{figure}

  When  $t$ is increased further, other interesting patterns appear. Varying time in steps of $10^{-4}$ units gives the  patterns in Fig. \ref{fig:varyT_10_4}, which  show that the wave function  and hence the probability pattern begin to spread.

It may be noted that  in all the above cases, the values of time  are irrational submultiples of $T=4/\pi$.  As observed in \cite{berry98,sanz19}, the patterns  in such cases have fractal structure. This can be seen  for cases with still larger values of time.  The following figures in  Fig. (\ref{fig:varyT_10_3}) (a)-(d) show the respective patterns, for values of $t=  2\times 10^{-3}$,  $t=  4\times 10^{-3}$,  $t=  6\times 10^{-3}$ and  $t=  8\times 10^{-3}$.

\begin{figure}[!b]
\resizebox {0.5 \textwidth} {0.25 \textheight }{\includegraphics {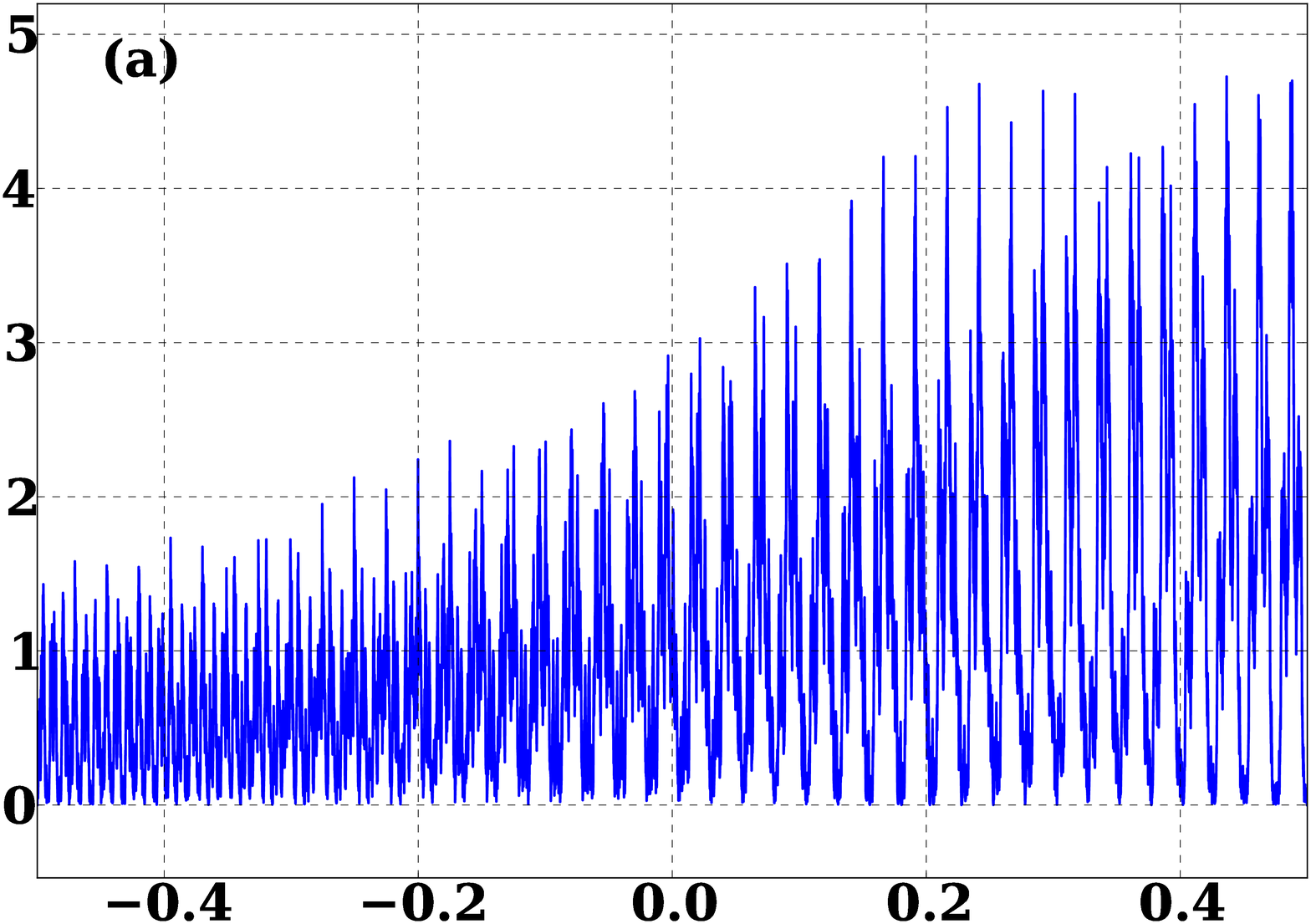}} ,\resizebox {0.5 \textwidth} {0.25 \textheight }{\includegraphics {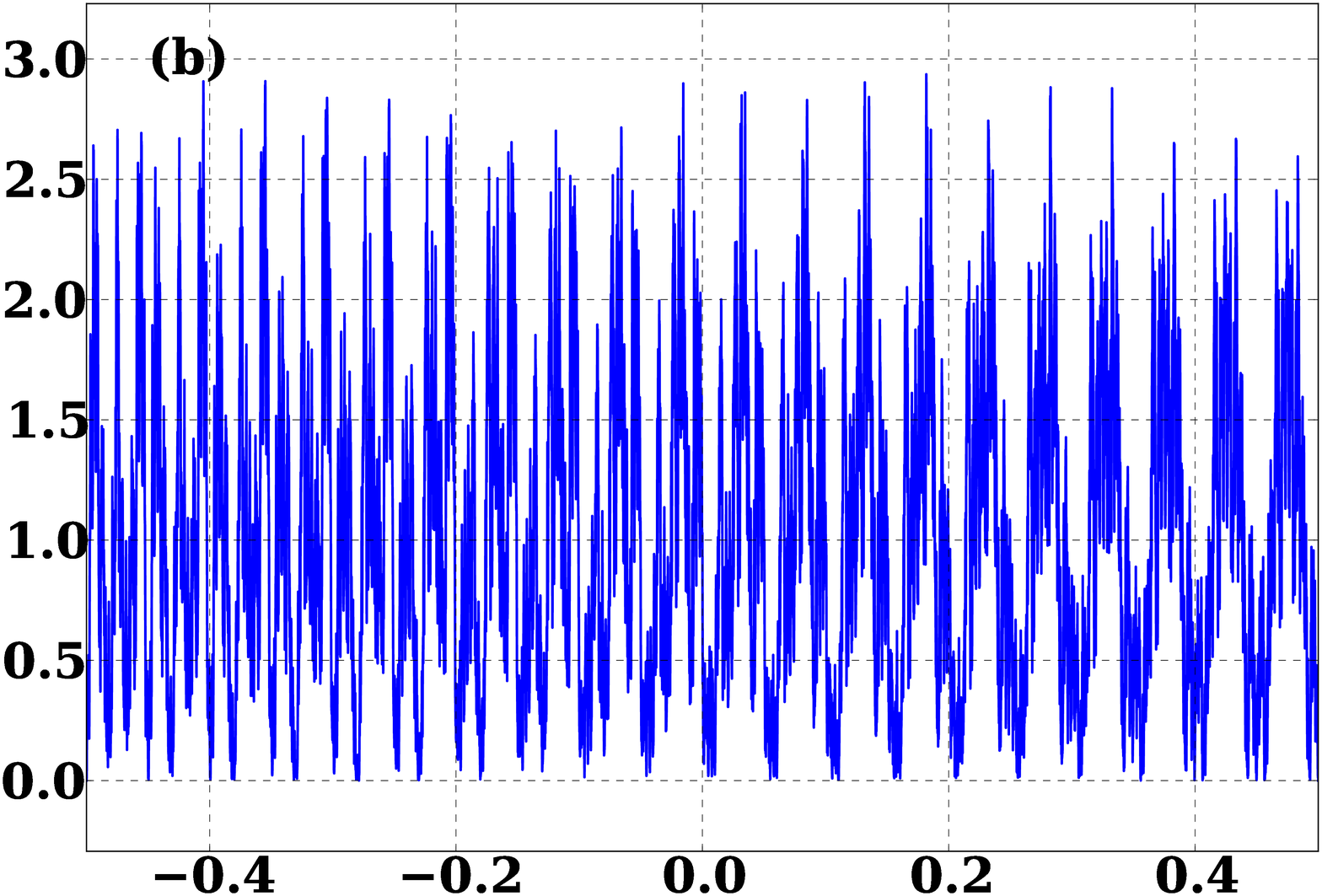}} ,\resizebox {0.5 \textwidth} {0.25 \textheight }{\includegraphics {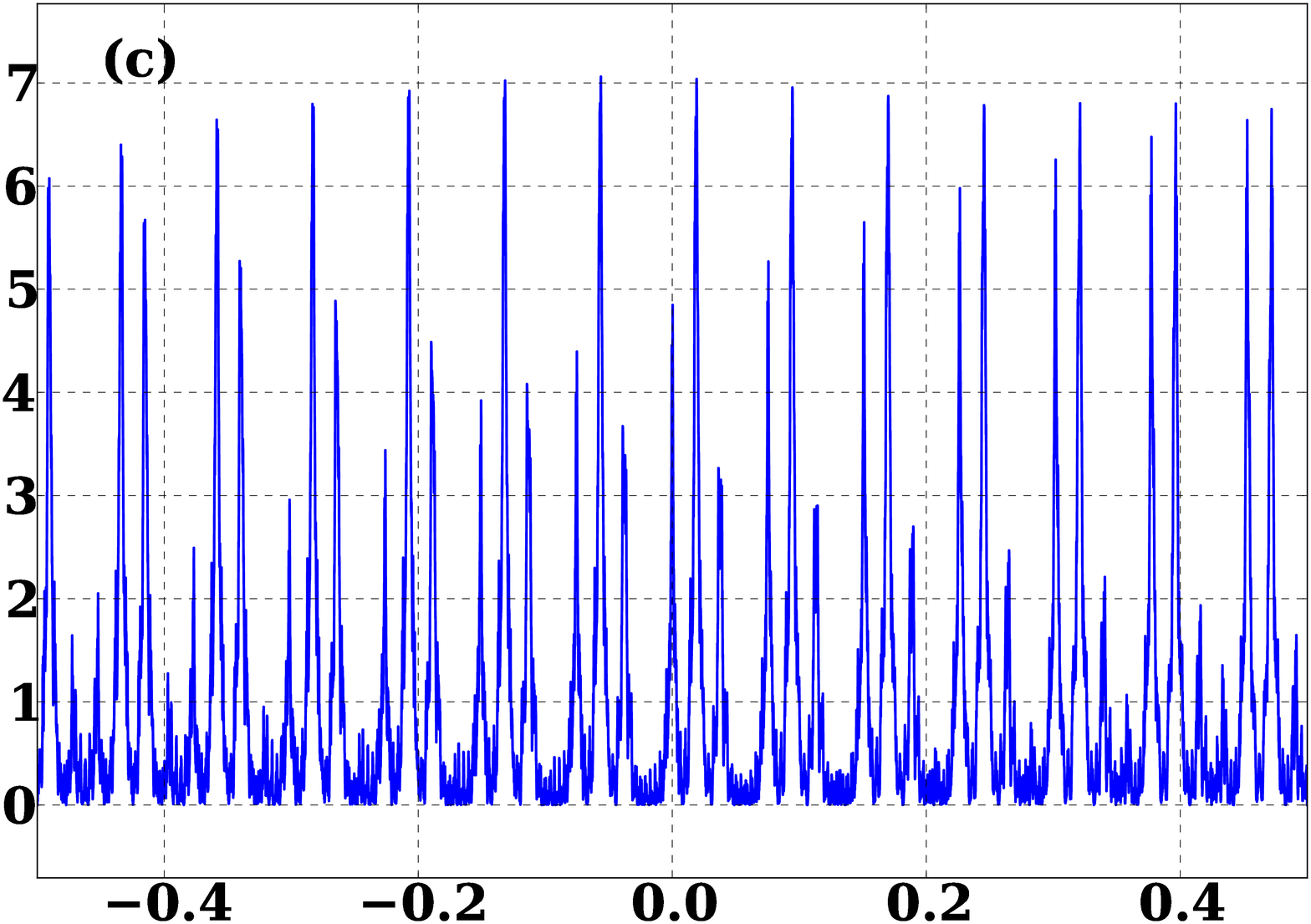}}  ,\resizebox {0.5 \textwidth} {0.25 \textheight }{\includegraphics {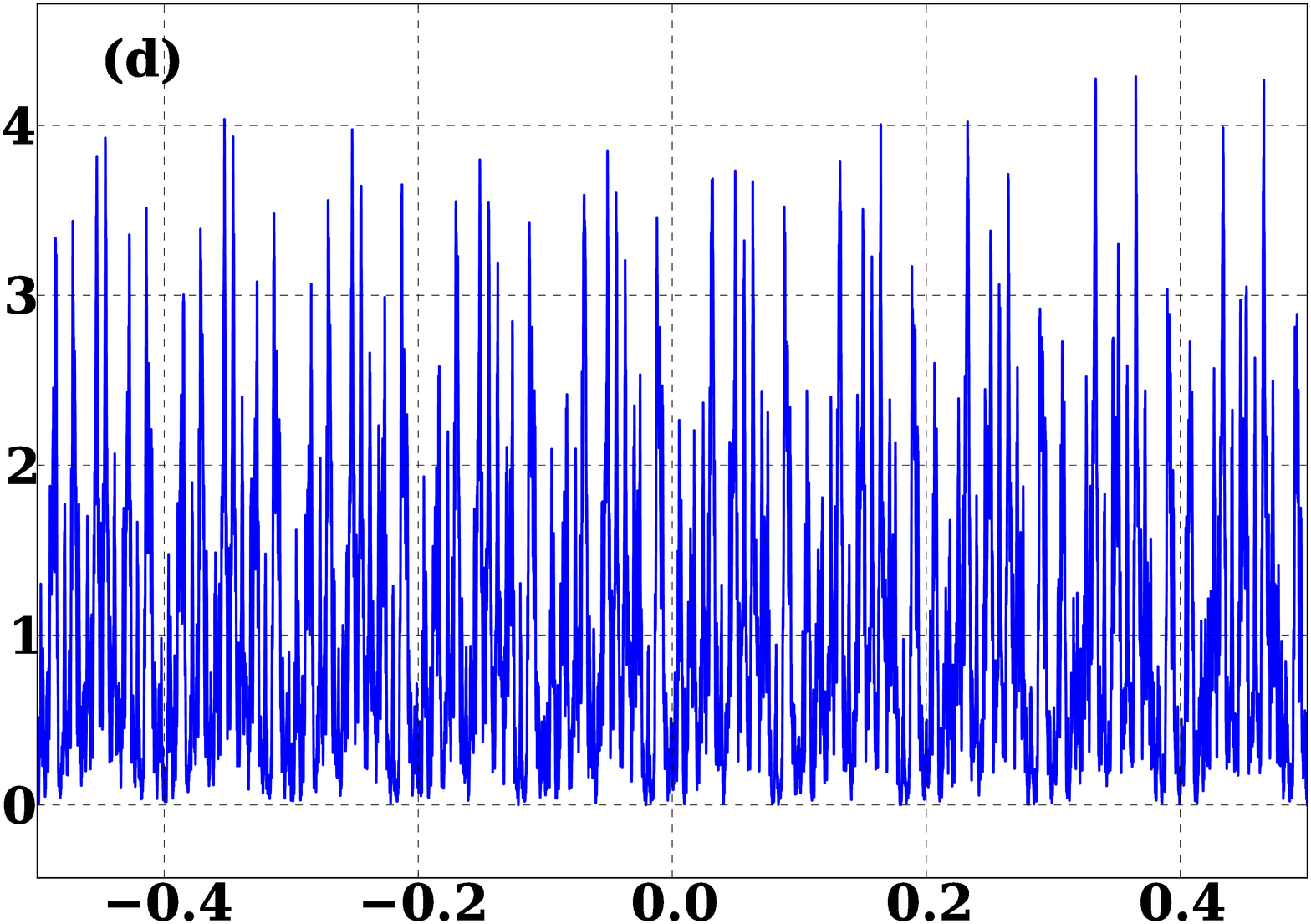}} 
\caption{ PMIC wave function plotted for time $N=50000$, with the values of (a) $t=2\times 10^{-3}$ (b)  $t=4\times 10^{-3}$, (c) $t=6\times 10^{-3}$ and $t=8\times 10^{-3}$} \label{fig:varyT_10_3}
\end{figure}

\begin{figure}[!b]
\resizebox {0.32 \textwidth} {0.25 \textheight }{\includegraphics {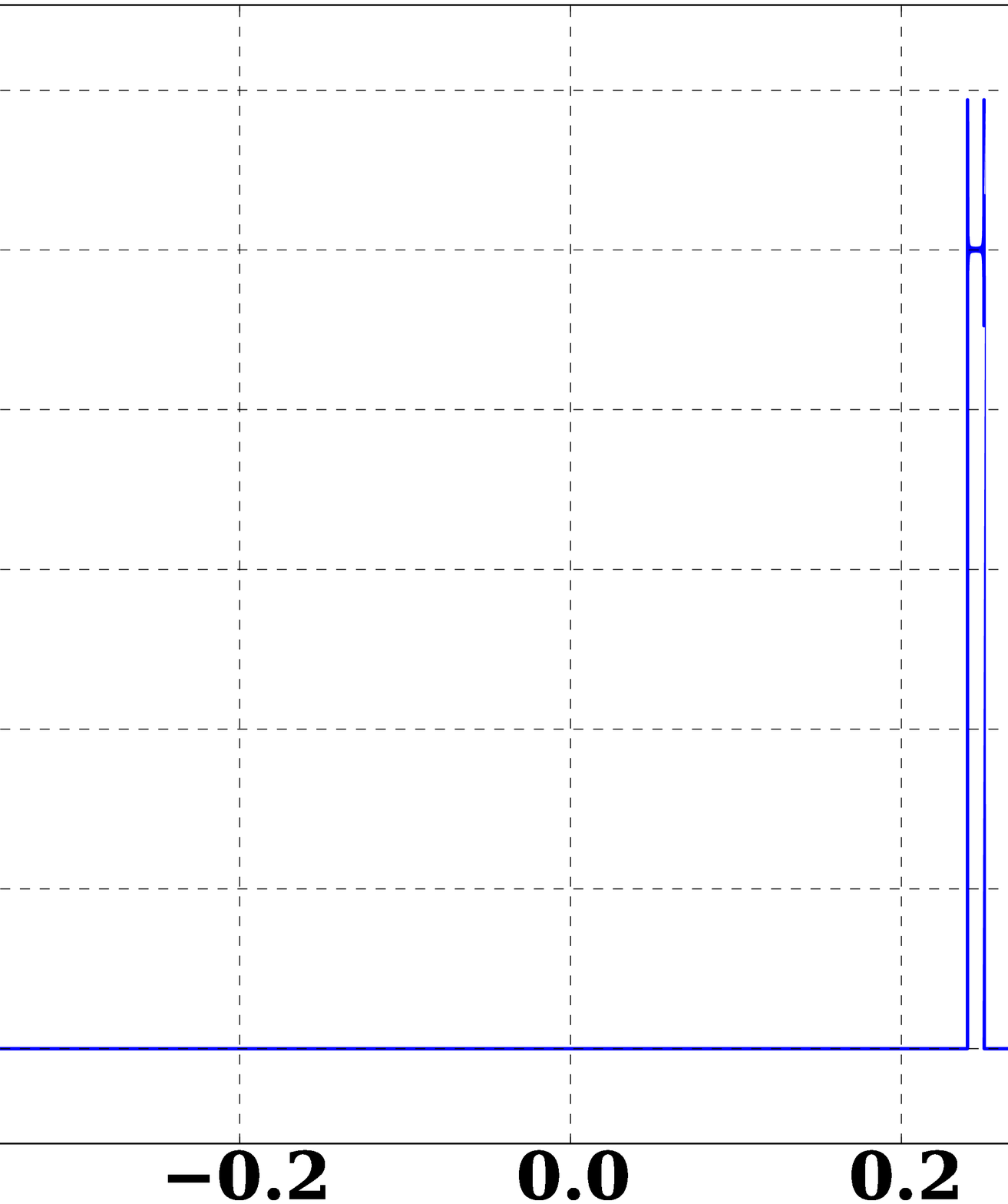}},\resizebox {0.32 \textwidth} {0.25 \textheight }{\includegraphics {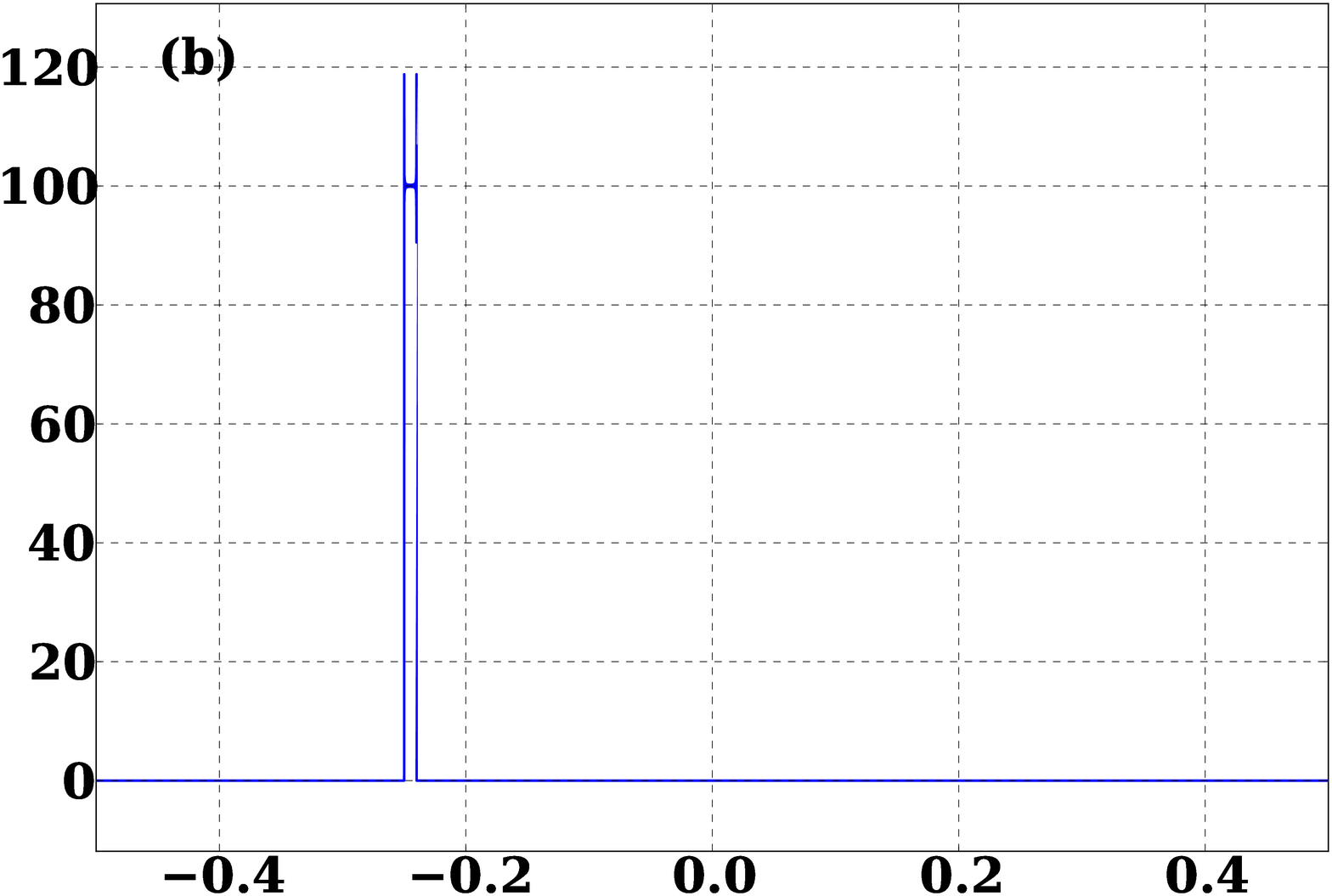}},\resizebox {0.32 \textwidth} {0.25 \textheight }{\includegraphics {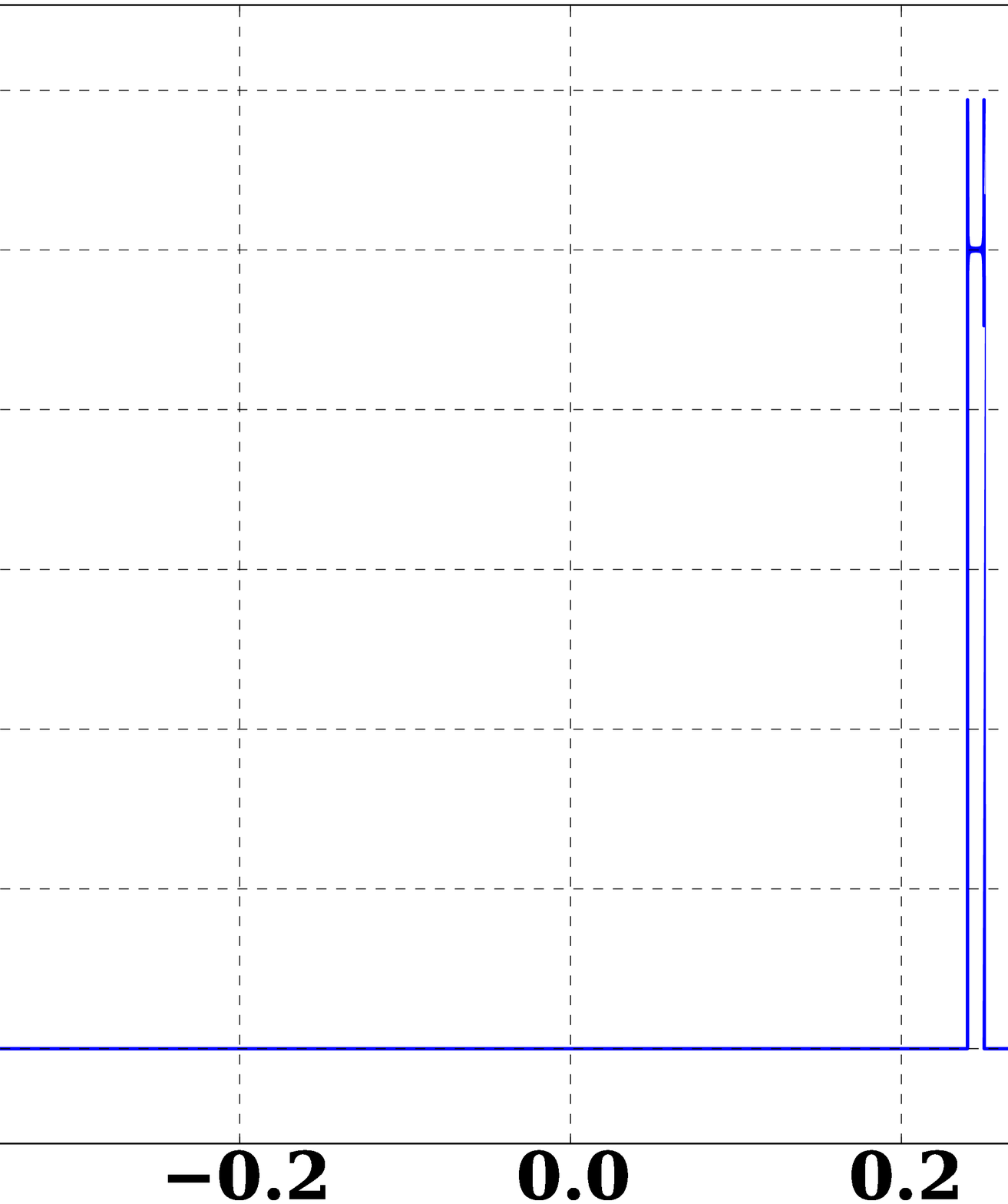}} 
\caption{ PMIC wave function plotted for time $N=50000$, with the values of (a) $t=0$ (b) $t=0.5  T$, (c) $t=T$}  \label{fig:varyT_half}
\end{figure}

Next  consider  cases where time $t$  is some rational submultiple of the revival time $T=4/\pi$.  Not surprisingly,  at half this time-period; i.e., at the time $t=T/2=2/\pi$, the rectangular wave  pattern is regained,  but with its  location  shifted to the opposite side with $y=-y_0$.  It  reappears at the same location $y=y_0$ at the end of the period $t=T$, as anticipated. Fig. \ref{fig:varyT_half} shows these patterns.

\begin{figure}[!b]
\resizebox {0.5 \textwidth} {0.25 \textheight }{\includegraphics {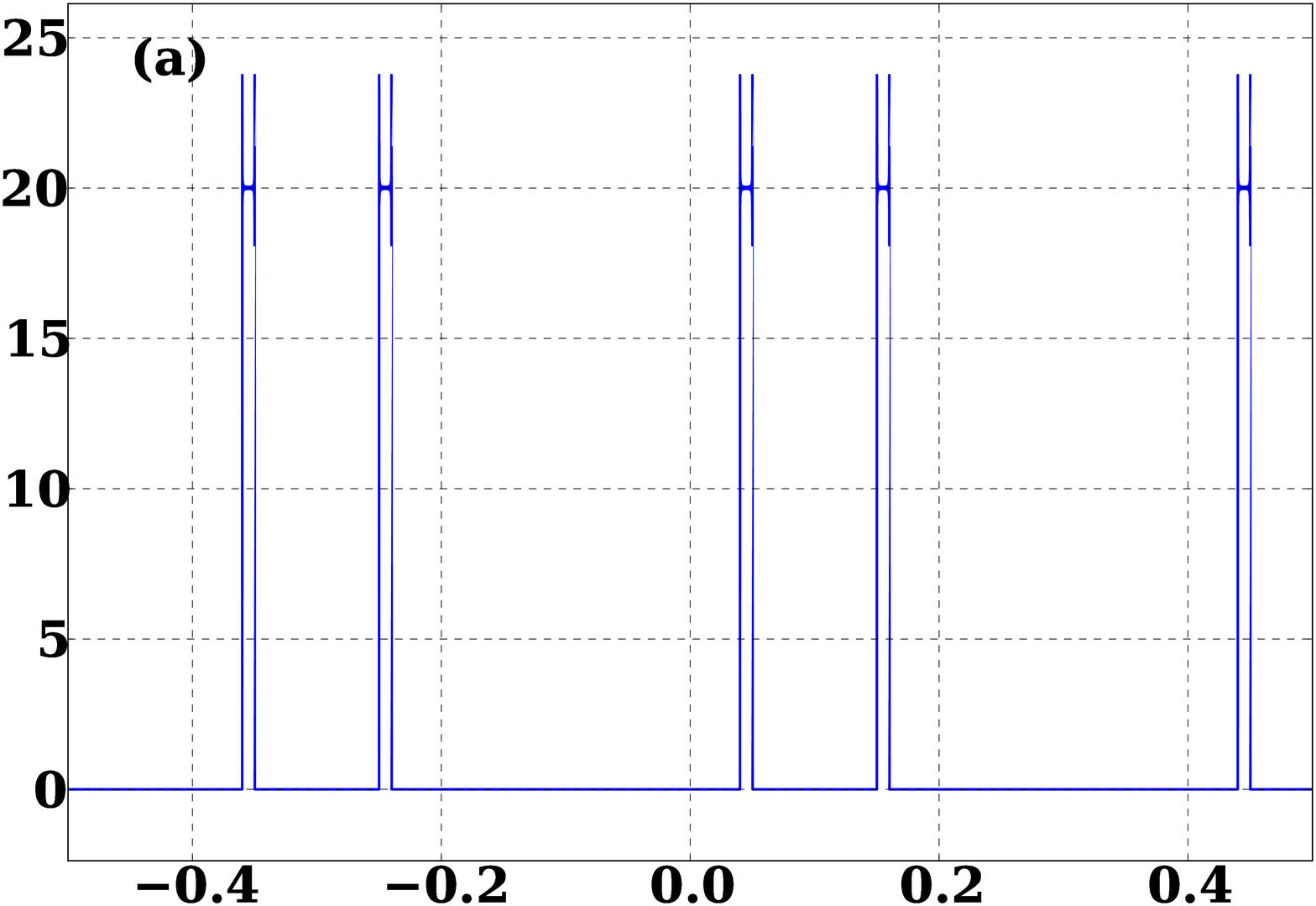}} ,\resizebox {0.5 \textwidth} {0.25 \textheight }{\includegraphics {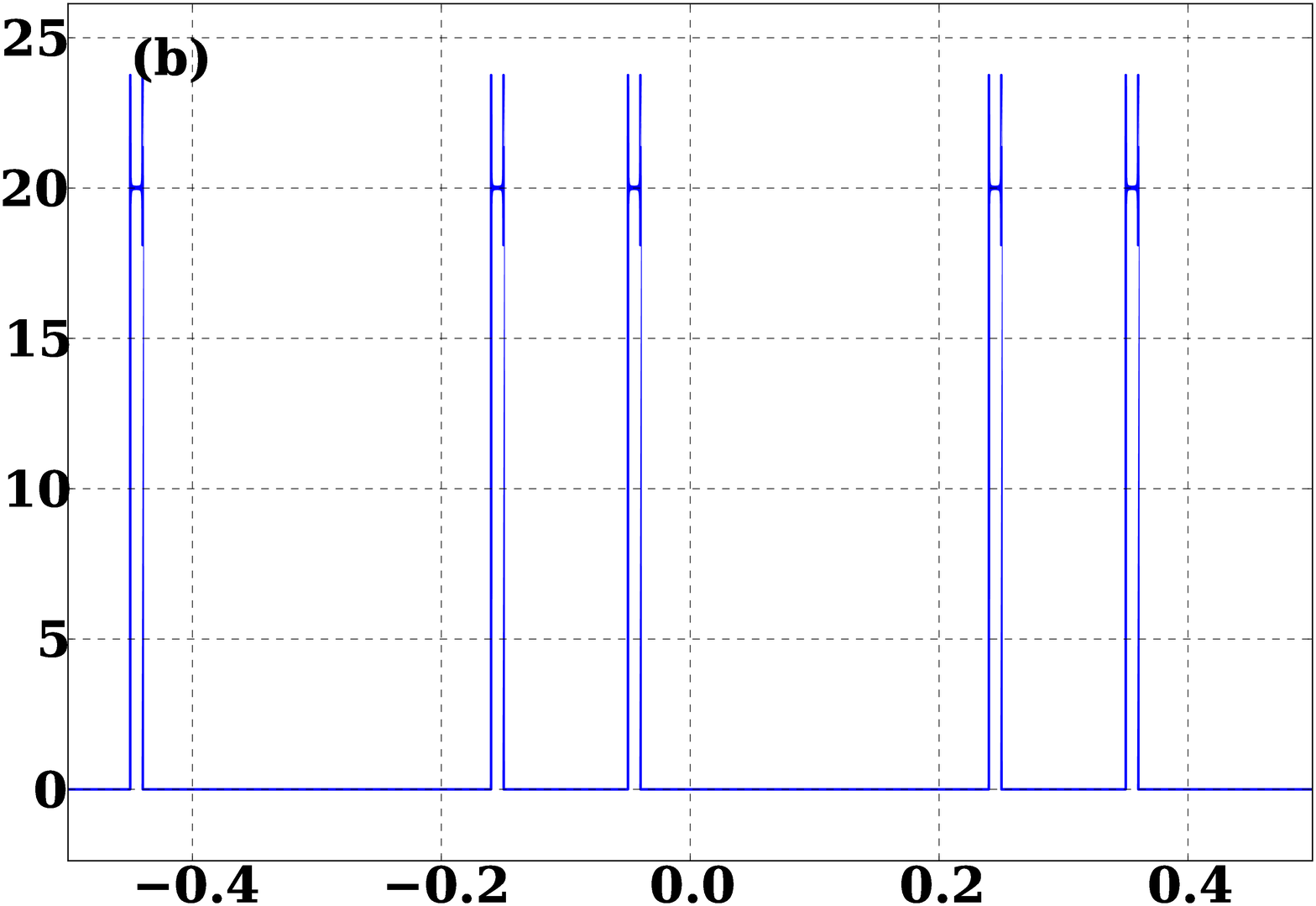}}, \resizebox {0.5 \textwidth} {0.25 \textheight }{\includegraphics {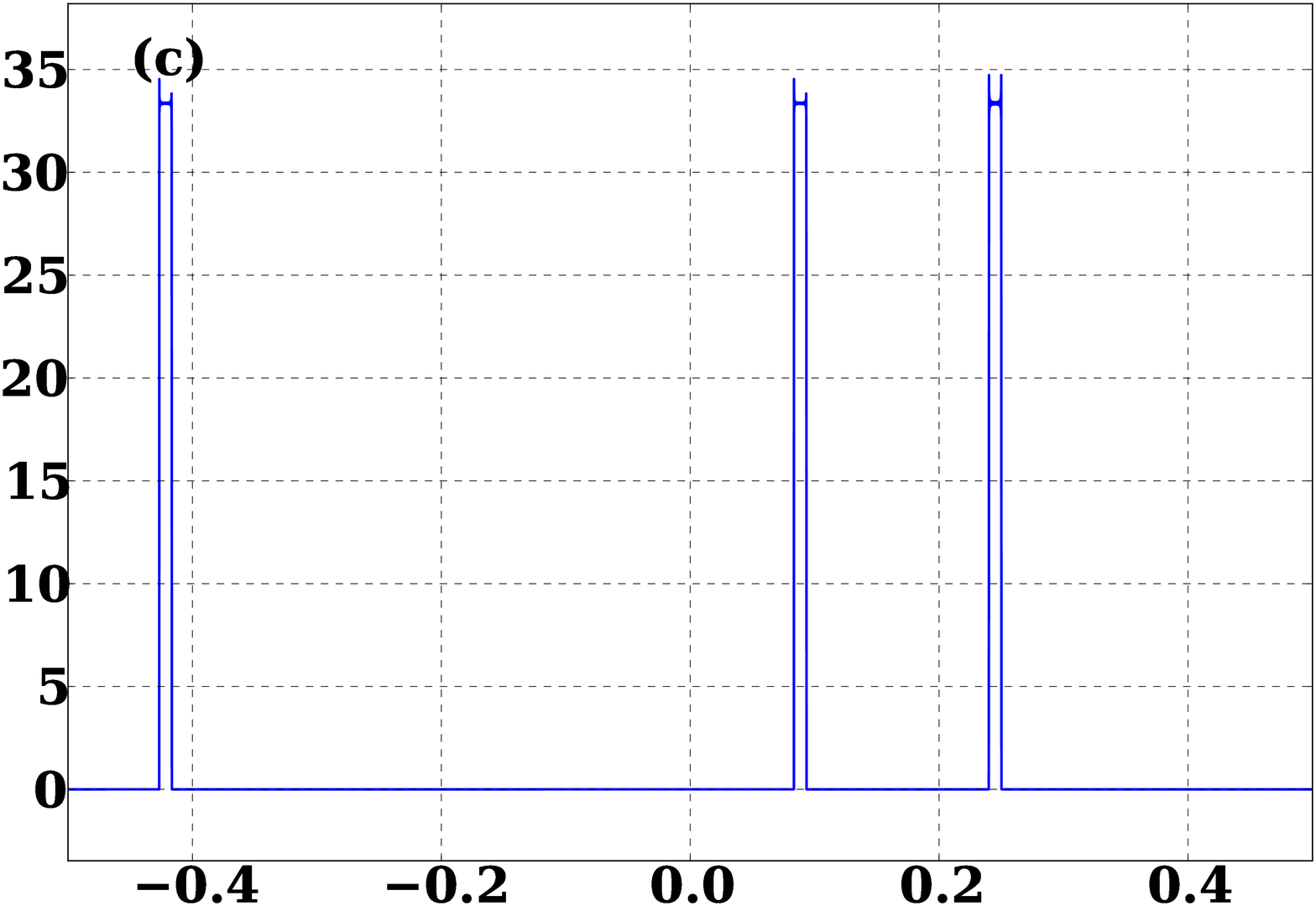}}, \resizebox {0.5 \textwidth} {0.25 \textheight }{\includegraphics {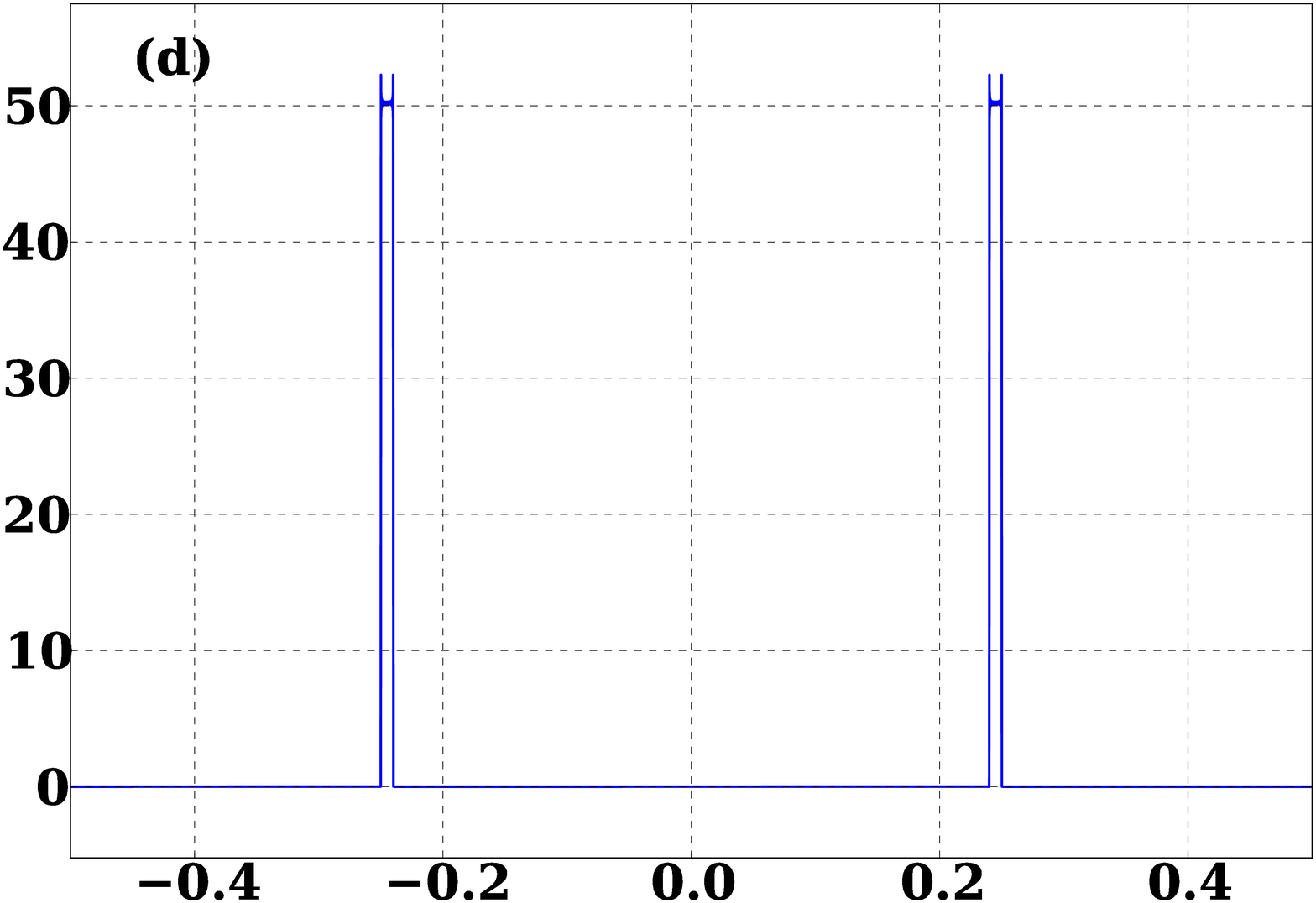}}  
\caption{ PMIC wave function plotted for time $N=50000$, with the values of (a) $t=0.1\;  T$, $t=0.3\;  T$, $t=0.7\;  T$ and $t=0.9\;  T$   (b)  $t=0.2\;  T$, $t=0.4\;  T$, $t=0.6\;  T$ and $t=0.8\;$ (c)  $t=T/3 \;$ (d)  $t=  T/4$}  \label{fig:varyT_evenodd}
\end{figure}

  One can observe the formation of  rectangular wave functions at other values of $t$ as well. For instance, at regular time intervals of $\Delta t=0.1\; T$, we have observed that  rectangular patterns appear. Some of these cases, where $t=0.1\;  T$, $t=0.3 \;T$, $t=0.7 \; T$ and $t=0.9\;  T$ show patterns as  given in Fig. \ref{fig:varyT_evenodd} (a) and in some other cases, where $t=0.2\;  T$, $t=0.4\;  T$, $t=0.6\;  T$ and $t=0.8\;  T$ the patterns are as  shown in Fig. \ref{fig:varyT_evenodd} (b). Similar calculations for other rational  fractions of $T$  are also shown, such as $t=T/3$ in Fig. \ref{fig:varyT_evenodd} (c), $t=T/4$ in Fig. \ref{fig:varyT_evenodd} (d).

 It was observed that the diffraction patterns are distinct  only for  $a\ll L$. As   the value of the slit-width $a$ is increased,  the Fraunhofer pattern appears somewhat late. But the revival time $T$ is independent of $a$. The most important feature to be noted is that the theoretically obtained patterns, especially those shown in Figs. \ref{fig:varyT_half} and \ref{fig:varyT_evenodd}, are testable in a diffraction experiment. The experiment proposed by us for testing these features  is described in the next section.

   \section{Experimental set up}

\begin{figure}[!b] 
  \resizebox {1 \textwidth} {0.4 \textheight }
{\includegraphics {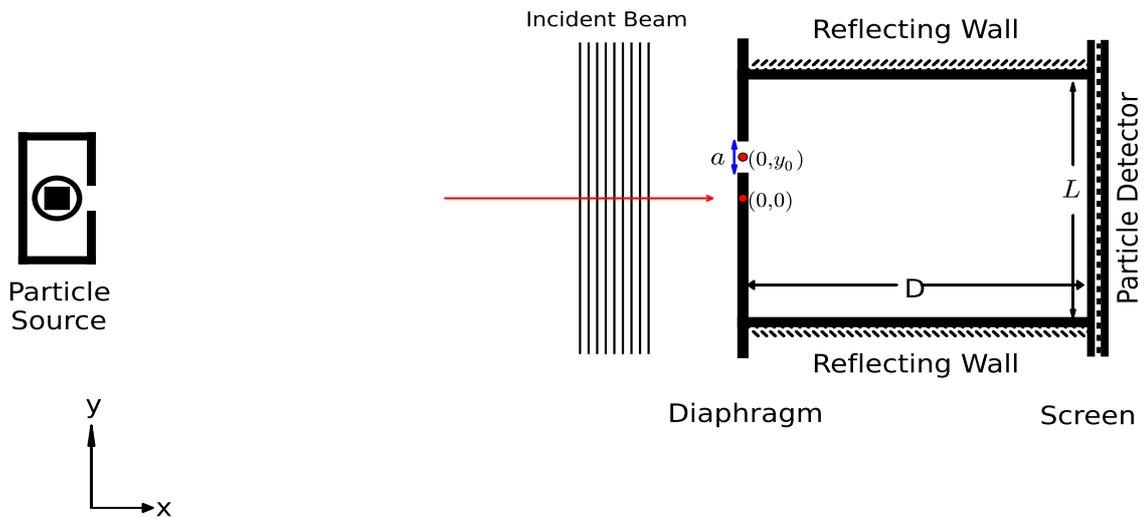}} 
\caption{Experimental set up }  \label{fig:apparatus}
  \end{figure}

   In the above, PMIC patterns were obtained theoretically  by following the standard axioms of quantum mechanics. Now we suggest a modified single-slit diffraction experiment  to test  these states. 
   
   The  arrangement consists of a source from which a monoenergetic, collimated particle beam propagates parallel to the $x$-axis, in the positive direction.   An impenetrable diaphragm of  small thickness, which can at the same time absorb the  particles hitting  it, is placed at $x=0$.  On the diaphragm, a slit of  finite width $\Delta y=a$ with center at $y=y_0$ and of infinite length along the $z$-axis  is cut. The  location of the slit is thus given by $x=0$, $y_0-a/2 \leq y\leq y_0+a/2$ and $-\infty< z <+\infty$.  In between the diaphragm and a screen placed at $x=D$, again the potential is zero. But let there be two  impenetrable and reflecting walls at $y=-L/2$ and $y=+L/2$, such that $y_0$ lies somewhere inside this interval. Thus a diffracted  particle that passed through the slit can be considered to be confined to an infinite potential well along the $y$-direction. Just like the diaphragm, the screen is completely absorbing. Additionally, it  serves to detect the particle, as in the case of experiments of diffraction and interference.     
   
   As per the geometry of the experimental set up,   the particles are free in the region with $x<0$ and hence can be described by a plane wave advancing in the positive $x$-direction.  This initial plane wave  collapses  when the particle passes  the diaphragm at $t=0$. Since   the particle that passed through the slit continues to be free along the $x$ and $z$-directions and is confined to an infinite potential well along the $y$-axis, the   collapse  is such that  the product wave function of the particle for $t>0$ and $x>0$ can  be written as

\begin{equation}
\Psi_x(x,t)\Psi_y^{y_0,a}(y,t)= \frac{N\; e^{i(k_xx-E_x t/\hbar)}}{ \sqrt{a}} \sum_n  \left[ \int_{y_0-a/2}^{y_0+a/2}dy^{\prime}\;u_n^{\star}(y^{\prime}) \right]u_n(y)e^{-iE_n t/\hbar}  . \label{eq:comb_wave_fn}
\end{equation}
Here $u_n$ and $E_n$ are given by Eqs. (\ref{eq:u_n}) and (\ref{eq:e_n}), respectively. The only affected component is the one along the the $y$-axis and it behaves as a PMIC wave function for $t>0$.  Since  $ \Psi_z(z,t)$  always remains a constant, we have omitted this part in the product wave function.

The probability distribution corresponding to this total wave function is independent of $x$ and $z$; it depends only on $y$ and $t$. The problem of obtaining a stationary diffraction pattern on the screen can be solved if we assume trajectories along the $x$-direction,  from the slit to the screen, as was done in \cite{mvjkm}. The postulate of existence of particle trajectories, in addition to the standard postulates of quantum mechanics,  is characteristic of nonlocal hidden variable theories such as the de Broglie-Bohm (dBB) \cite{db1}, modified de Broglie-Bohm (MdBB) \cite{mvj1}, and the Floyd-Faraggi-Matone (FFM) \cite{floyd,faraggi,carroll} trajectory formulations.  In the present case, the  $x$-component of velocity has the same value $v_x=\hbar k_x/m$   in all these three formalisms. Therefore we have taken the time  with which a particle from the slit at $x=0$ reaches the screen placed at $x=D$  as $t=D/v_x$.
   Using this time $ t$ in Eq. (\ref{eq:comb_wave_fn}), one can plot $|\Psi_x(x,t)\Psi_y^{y_0,a}(y,t)|^2$ against $y$ on the screen placed at a distance $D$. Actually, these are the same patterns with  $|\Psi_y^{y_0,a}(y,t)|^2$ plotted against $y$ in Fig. (1)-(7) discussed in Sec. {\ref{sec:inf_pot_well}}, where $t=Dm/(\hbar k_x)$. Predictions under this scheme are without any adjustable parameters and hence exact. In an actual experiment,  obtaining those special patterns given in Fig. (2)-(5) may be possible with very good accuracy. Of particular interest is  the occurrence of Fresnel and Fraunhofer patterns in this case.  
Moreover, it may be  noted that since there exists a revival time $T$ for the wave function,  the rectangular  pattern  will reappear at regular intervals as we vary $D$. The distance corresponding to the  period with which it repeats is $D_T=\frac{4 \hbar k_x}{\pi m}$. This is another clearly verifiable  prediction of the theory.

\section{Discussion}
Here we treated the passing of the particle through a single-slit  as the measurement of its position. A second measurement of the particle  happens at the detector or screen, where it is completely absorbed. Our attempt was to study the time-evolution of the state of the system in between these two measurements. For that, we assumed that as a result of the first measurement,  the plane wave function of the incident particle collapsed to a rectangular function, with a discontinuity at the edges of the slit. The attempt in \cite{marcella} was  to  go to the momentum space and evaluate the probability distribution for the momentum variable. It is then tacitly assumed that the particles move towards the screen with the corresponding momentum values \cite{rothman}, as it would do in the case of classical mechanics. By this approach, only the Fraunhofer-type patterns were obtained on the screen, for all  distances from the slit. In contrast, in \cite{mvjkm}  we postulated that the time-evolution of  the collapsed state is described by the PMIC wave function.  It was shown that this results in a single expression that describes the Fresnel and Fraunhofer diffraction. 

In the present work, what we suggest  is an experiment to test the PMIC state in a more precise way. The  apparatus  in our  experiment is a modified version of  Lloyd's mirror in  optics. The standard Lloyd's mirror arrangement, which is originally designed to show interference \cite{born_wolf}, has only a primary  source and its virtual image and no slits. In this form, the   experiment is simpler than Young's double-slit experiment, since the latter has the diffraction pattern due to two slits overlaid by an interference between the two sources. The Lloyd's mirror can have only interference of light from two sources, and  no  slits. But our case discussed in this paper  is similar to having two Lloyd's mirrors, with a primary slit and an infinite number of virtual slits as sources.  The repetition of rectangular pattern with the variation of $D$,  as predicted  by us using the PMIC formalism, is not anticipated in a wave optics treatment of this experiment. The verification of PMIC states by this experiment will be a step towards better understanding of `quantum collapse' in general.

An important point that  deserves serious attention is the value of $N$  used in the series expansion in Eqs. (\ref{eq:slit_wavefn2}), (\ref{eq:slit_wavefn3}) or (\ref{eq:comb_wave_fn}). In practice, while  plotting the wave function in the form of a converging infinite series, a finite upper limit for $n$ is unavoidable. We have put this upper limit in our calculations to be   $N=50000$, so that a nearly rectangular function at $t=t_M$ is obtained. Any  function can be written as a superposition of Dirac delta functions and it is well known that a delta function will be exact only when the limit $N\rightarrow \infty$ is taken in its closure representation. A problem related to the eigenstate of position  is that the expectation value of energy  shall tend to infinity.  Consequently, the measurement of position with infinite precision is impossible.  This is also  true for wave functions with sharp discontinuities, such as the  perfect rectangular wave function considered by us.  A non-zero constant wave  function inside the box, as considered in \cite{berry98}, requires infinite energy.  However,  physical objects such as the slit in this experiment will not have arbitrarily sharp boundaries and hence it is justifiable to fix a finite and large $N$ that corresponds to a moderately round edge.  The energy-time uncertainty relation can also be invoked to justify a finite but very large value for $N$.  In the  experiment such as the diffraction through a single slit, the uncertainty in energy can be  large but finite during the time of  transit of the particle through the slit, which  can be  very small but nonzero. (The slit is assumed to be made on a thin diaphragm.) Moreover, since in this paper we are more concerned with the diffraction patterns, an upper limit for $N$ does not pose much problem because,  as we have  verified in Fig. \ref{fig:varyN},  the shape of the curves have no appreciable change on further increasing the value of $N$.    In particular, it may be noted   that our main  predictions, such as the Fraunhofer and Fresnel patterns, the reappearance of the rectangular patterns at particular values of distance $D$, etc.  will remain unaffected even if we fix a finite, sufficiently large value for $N$.
   
In this paper  we have also used  the trajectory representation in quantum mechanics, which  accept the existence of   particle trajectories  associated with the wave functions. In \cite{mvjkm}, we have used this additional postulate to obtain the single expression that gives a unified description of Fresnel and Fraunhofer diffractions. It was noted that all the three trajectory formulations make identical predictions in this case and that the success of our PMIC description supports  the existence of particle trajectories. If verified, the predictions in this paper  will also   support the existence of  particle trajectories.

\label{lastpage}
\end{document}